\documentclass[10pt,a4paper,twocolumn]{IEEEtran}

\usepackage{amssymb}
\usepackage{graphicx}
\usepackage{multirow}
\usepackage{float}
\usepackage{amsmath,amssymb,amsfonts,mathrsfs}
\usepackage{color}
\usepackage{subfig}

\newtheorem{theorem}{\textbf{Theorem}}
\newtheorem{lemma}{\textbf{Lemma}}
\newtheorem{corollary}{\textbf{Corollary}}

\newtheorem{definition}{\textbf{Definition}}

\begin{document}
\renewcommand{\thepage}{18--\arabic{page}}
\setcounter{page}{1}
\setlength{\topsep}{0.0cm}
\setlength{\itemsep}{0.0em}

\title{\Large\bf Structure Identifiability of an NDS with LFT Parametrized Subsystems} 

\author{Tong Zhou 
\thanks{This work was supported in part by the NNSFC under Grant 61733008,  52061635102 and 61573209.}
\thanks{Tong Zhou is with the Department of Automation, Tsinghua University, Beijing, 100084, P.~R.~China
        {(email: {\tt\small tzhou@mail.tsinghua.edu.cn).}}%
}}
\maketitle

\begin{abstract}                          
Requirements on subsystems have been made clear in this paper for a linear time invariant (LTI) networked dynamic system (NDS), under which subsystem interconnections can be estimated from external output measurements. In this NDS, subsystems may have distinctive dynamics, and subsystem interconnections are arbitrary. It is assumed that system matrices of each subsystem depend on its (pseudo) first principle parameters (FPPs) through a linear fractional transformation (LFT). It has been proven that if in each subsystem, the transfer function matrix (TFM) from its internal inputs to its external outputs is of full normal column rank (FNCR), while the TFM from its external inputs to its internal outputs is of full normal row rank (FNRR), then \textcolor{black}{the structure of the NDS is identifiable}. Moreover, under some particular situations like there is no direct information transmission from an internal input to an internal output in each subsystem, a necessary and sufficient condition is established for NDS structure identifiability. A matrix valued polynomial (MVP) rank based equivalent condition is further derived, which depends affinely on subsystem (pseudo) FPPs and can be independently verified for each subsystem. From this condition, some necessary conditions are obtained for both subsystem dynamics and its (pseudo) FPPs, using the Kronecker canonical form (KCF) of a matrix pencil.
\end{abstract}


\section{Introduction}
\setlength{\itemsep}{0.0em}

Networked dynamic systems (NDS) have been attracting research attentions for a long time, which are some times also called large scale systems, composite systems, especially in the 60s of last century \cite{Siljak1978,sbkkmpr2011,zyl2018}. With technology developments, especially those in communications and computers, the scale of an NDS becomes larger and larger. Moreover, some new issues also arise, such as attack prevention, random communication delay/failure, etc. On the other hand, some classic problems including revealing the structure of an NDS from measurements, computationally efficient conditions for NDS controllability/observability verifications, etc., still remain challenging \cite{cpakj2017,cw2020,htw2009,pka2016,pmssaxcas2010,scl2015,xz2014}. Among these, an essential issue is NDS identification which is widely realized as the basis for developing effective methods in NDS analysis and synthesis \cite{cw2020,vdhb2013, zyl2018}.

Particularly, in order to monitor the behaviors of an NDS or to improve its performances, it is usually required to understand the dynamics of its subsystems, as well as their interactions. While in some applications both of them are known from the NDS working principles and/or constructions, there are also various situations in which both of them or one of them must be estimated from experiment data. For example, in an NDS with wireless communications, some subsystem interactions may fail to work due to unpredictable communication congestion; in an NDS constituted from several mobile robots, information exchange among these robots may vary with changing environments; in a gene regulation network, a direct interaction is usually hard and/or too expensive to measure; etc. In these applications, it is invaluable to understand subsystem interactions from measured experiment data \cite{cw2020,pmssaxcas2010,sbkkmpr2011,vdhb2013, zyl2018}.

In NDS dynamics description, the adopted approaches can be briefly divided into two categories. One of them treat each measured variable as a node, while transfer functions among these variables as edges, which has been used in many researches on NDS analysis and synthesis. Examples includes \cite{cw2020,hgb2019,wvd2018} and the references therein. The other approach treats each subsystem as a node, while interactions among subsystems as edges. This approach has also been widely adopted and appears more natural and popular, for example \cite{cw2017,cw2020,pmssaxcas2010,Siljak1978,sbkkmpr2011,vtc2021,zyl2018} and the associated references. \textcolor{black}{In \cite{cw2020}, characteristics of these two descriptions are investigated from a structural informativity aspect, comparing with the extensively adopted transfer function matrix (TFM) description and the state space model. An important observation there is that while all these four descriptions can describe the same input/output behavior of a system, each one may characterize a different notion of system structure. In addition, no matter which of the aforementioned two approaches is adopted in NDS dynamics descriptions, further efforts are still required for developing efficient methods that estimate the associated model from experiment data. The latter has also been observed in \cite{wvd2018,vtc2021,zyl2018} and the references therein.}

In particular, several recent studies make it clear that there is no guarantee that NDS subsystem interactions can always be estimated from experiment data. For example, it is shown in \cite{cw2020} that even if the TFM of an NDS can be perfectly estimated, there are still possibilities that its subsystem interactions can not be identified. \textcolor{black}{More specifically, the subsystem structure description is in general more structurally informative than a TFM description.} To clarify situations under which NDS structure can be identified, some eigenvector based conditions are derived in \cite{ptt2019} for an NDS with descriptor subsystems and diffusive subsystem coupling, so that variations of its subsystem interactions can be detected. \cite{vtc2021} studies topology identifiability when subsystems of an NDS are coupled through their outputs. It is proven that an NDS is topologically identifiable only when the constant kernel of a TFM which is completely determined by subsystem dynamics, is equal to a zero vector. It has also been shown there that this condition becomes also sufficient under some special situations.

In this paper, we investigate requirements on the NDS model adopted in \cite{Zhou2015,zyl2018,zz2020} such that its subsystem interactions can be identified from experiment data. In this NDS, each subsystem is permitted to have distinctive dynamics, and subsystem interactions are directed. In addition, the system matrices of each subsystem depend on its (pseudo) first principle parameters (FPP) through a linear fractional transformation (LFT). This NDS model includes those adopted in \cite{cw2020,ptt2019,vtc2021} as special cases, and may be considered as the most general one among linear time invariant NDS models. It is proven that \textcolor{black}{the structure of this NDS model is identifiable}, if two transfer function matrices (TFM) associated with each of its subsystems independently, are respectively of full normal column rank (FNCR) and full normal row rank (FNRR). Based on this result, it is further shown that this FNCR/FNRR condition is respectively equivalent to the FNCR of two matrix valued polynomials (MVP) that depends affinely on the (pseudo) FPPs of each subsystem. \textcolor{black}{Moreover, a necessary and sufficient condition is established for NDS structure identifiability, under the following two situations. The first is that in each subsystem, there is no direct information transmission from an internal input to an internal output. The second is that for each subsystem, the TFM from its internal inputs to its external outputs can be expressed as the multiplication of a FNCR TFM and the TFM from its internal inputs to its internal outputs, while the TFM from its external inputs to its internal outputs can be expressed as the multiplication of the TFM from its internal inputs to its internal outputs and a FNRR TFM.} This condition can be verified for each \textcolor{black}{tuple of} subsystems independently, and its computational complexity increases only quadratically with the number of subsystems in an NDS.

The outline of the remaining of this paper is as follows. At first, in Section II, problem descriptions are given, together with the NDS model adopted in this paper and some preliminary results. NDS structure identifiability is studied in Section III, in which some necessary and sufficient conditions on the transfer function matrices of a subsystem are derived. Section IV investigates relations between NDS structure identifiability and subsystem (pseudo) FPPs. A numerical example is given in Section V to illustrate applicability of the theoretical results obtained in Sections III and IV. Some concluding remarks are given in Section VI in which several further issues are discussed. Finally, an appendix is included to give proofs of some technical results.

The following notation and symbols are adopted. ${\cal C}$ and ${\cal C}^{m}$ stand respectively for the set of complex numbers and the $m$ dimensional complex Euclidean space. ${\rm\bf det}\left(\cdot\right)$ represents the determinant of a square matrix, ${\rm\bf null}\left(\cdot\right)$ the (right) null space of a matrix, $\cdot^{\perp}$ the matrix whose columns form a base of the (right) null space of a matrix, while $A\otimes B$ the Kronecker product of two matrices. ${\rm\bf diag}\{X_{i}|_{i=1}^{L}\}$ denotes a block diagonal matrix with its $i$-th diagonal block being $X_{i}$, while ${\rm\bf col}\{X_{i}|_{i=1}^{L}\}$ the vector/matrix stacked by $X_{i}|_{i=1}^{L}$ with its $i$-th row block vector/matrix being $X_{i}$, and ${\rm\bf vec}\{X\}$ the vector stacked by the columns of the matrix $X$. $I_{n}$, $0_{m}$ and $0_{m\times n}$ represent respectively the $m$ dimensional identity matrix, the $m$ dimensional zero column vector and the $m\times n$ dimensional zero matrix. The subscript is usually omitted if it does not lead to confusions. The superscript $T$ is used to denote the transpose of a matrix/vector.

\section{Problem Description and Some Preliminaries}

In a real world NDS, its subsystems may have distinctive dynamics. When an NDS is linear and time invariant (LTI), a model is suggested in \cite{Zhou2015,zyl2018,zz2020} to describe relations among subsystem inputs, outputs and its (pseudo) FPPs. More specifically, for an NDS $\bf \Sigma$ consisting of $N$ subsystems, the following model is utilized to describe the dynamics of its $i$-th subsystem ${\bf{\Sigma}}_i$.
\begin{equation}
\left[\!\! {\begin{array}{c}
{\delta({{x}}(t,i))}\\
{{z}(t,i)}\\
{{y}(t,i)}
\end{array}}\!\! \right]  \!\!=\!\!
\left[\!\! \begin{array}{ccc}
{A_{\rm\bf xx}(i)} & {A_{\rm\bf xv}(i)} & {B_{\rm\bf x}(i)}\\
{A_{\rm\bf zx}(i)} & {A_{\rm\bf zv}(i)} & {B_{\rm\bf z}(i)}\\
{C_{\rm\bf x}(i)} & {C_{\rm\bf v}(i)} & {D_{\rm\bf u}(i)}
\end{array}\!\! \right]\!\! \left[\!\! \begin{array}{c}
{{x}(t,i)}\\
{{v}(t,i)}\\
{{u}(t,i)}
\end{array}\!\! \right]
\label{eqn:1}
\end{equation}
Moreover, interactions among NDS subsystems are described by the following equation
\begin{equation}
v(t)=\Phi z(t)
\label{eqn:2}
\end{equation}
in which $z(t)$ and $v(t)$ are assembly expressions respectively for the internal output vectors and the internal input vectors of the whole NDS. That is, $z(t)={{\rm{{\bf{{\rm {col}}}}}}\{z(t,i)|_{i=1}^N\}}$ and $v(t)={{\rm{{\bf{{\rm {col}}}}}}}\{v(t,i)|_{i=1}^N\}$ respectively. In addition, the system matrices $A_{\rm\bf xx}(i)$ etc. of the subsystem ${\bf{\Sigma}}_i$ are assumed to depend on its (pseudo) FPPs through the following LFT,
\begin{eqnarray}
& &\hspace*{-1.0cm} \left[\!\!\! \begin{array}{ccc}
{A_{\rm\bf xx}(i)} & {A_{\rm\bf xv}(i)} & {B_{\rm\bf x}(i)}\\
{A_{\rm\bf zx}(i)} & {A_{\rm\bf zv}(i)} & {B_{\rm\bf z}(i)}\\
{C_{\rm\bf x}(i)} & {C_{\rm\bf v}(i)} & {D_{\rm\bf u}(i)}
\end{array}\!\!\! \right] \!\!=\!\!
\left[\!\!\! {\begin{array}{ccc}
{A_{\rm\bf xx}^{[0]}(i)} & {A_{\rm\bf xv}^{[0]}(i)} & {B_{\rm\bf x}^{[0]}(i)}\\
{A_{\rm\bf zx}^{[0]}(i)} & {A_{\rm\bf zv}^{[0]}(i)} & {B_{\rm\bf z}^{[0]}(i)}\\
{C_{\rm\bf x}^{[0]}(i)} & {C_{\rm\bf v}^{[0]}(i)} & {D_{\rm\bf u}^{[0]}(i)}
\end{array}}\!\!\! \right] \!\!+  \nonumber\\
& & \hspace*{-0.80cm}\left[ \!\!{\begin{array}{l}
{H_{\rm\bf x}(i)}\\
{H_{\rm\bf z}(i)}\\
{H_{\rm\bf y}(i)}
\end{array}} \!\!\right]\!\! P(i)\left[ I_{m_{{\rm\bf g}i}} - {G{(i)}}{P{(i)}}\right]^{\!-1}\!\left[\!\! \!\!\begin{array}{ccc}
F_{\rm\bf x}(i)\! & \!F_{\rm\bf v}(i)\! & \!F_{\rm\bf u}(i)
\end{array}\!\! \right]
\label{eqn:1-a}
\end{eqnarray}
in which the matrix $P(i)$ is in principle constituted from fixed zero elements and (pseudo) FPPs of Subsystem ${\bf{\Sigma}}_i$, while all the other matrices are prescribed. Moreover, $m_{{\rm\bf g}i}$ stands for the number of the rows of the matrix $G(i)$.

A (pseudo) FPP may be a concentration or a reaction ratio in biological/chemical processes, a resistor, an inductor or a capacitor in electrical/electronic systems, a mass, a spring or a damper in mechanical systems, etc., or a simple function of them, which can usually be chosen/tuned in system designs. The matrices $G{(i)}$, $H_{\rm\bf \star}{(i)}$ with $\star={ \rm\bf x}$, $\rm\bf z$ or $\rm\bf y$, and $F_{\rm\bf \star}{(i)}$ with $\star={\rm\bf x}$, $\rm\bf v$ or $\rm\bf u$, are introduced to indicate how the system matrices of this subsystem is changed by its (pseudo) FPPs. These matrices, as well as the matrices
${A_{\rm\bf *\#}^{[0]}(i)}$, ${B_{\rm\bf *}^{[0]}(i)}$, ${C_{\rm\bf *}^{[0]}(i)}$ and ${D_{\rm\bf u}^{[0]}(i)}$, in which ${\rm\bf *,\#}={\rm\bf x}$, ${\rm\bf u}$, ${\rm\bf v}$, ${\rm\bf y}$ or ${\rm\bf z}$, are often used to
represent chemical, biological, physical or electrical principles governing subsystem dynamics, such as Newton's mechanics, the Kirchhoff's current law, etc., which implies that they are usually prescribed and can hardly be chosen or tuned in designing a system.

In the above NDS model, $\delta(\cdot)$ denotes either the derivative of a function with respect to time or a forward time shift operation. In other words, the above model can be either continuous time or discrete time. Moreover, $t$ stands for the temporal variable, $x(t,i)$ the state vector of its $i$-th subsystem ${\bf{\Sigma}}_i$, $y(t,i)$ and $u(t,i)$ its external output and input vectors respectively,  $z(t,i)$ and $v(t,i)$ its internal output and input vectors respectively, representing signals sent to other subsystems and signals gotten from other subsystems. In addition, the matrix $\Phi$ describes influences among different NDS subsystems, and is called subsystem connection matrix (SCM). If each subsystem  is regarded as a node and each nonzero element of its SCM as an edge, a graph can be associated with an NDS, which is usually called the structure or topology of the corresponding NDS.

The above model reflects the well known fact that in a real world plant, elements in its system matrices are usually not algebraically independent of each other, and some of them can even not be tuned in system designs. A more detailed discussion can be found in \cite{Zhou2015,zyl2018,zz2020} on engineering motivations of the aforementioned model. To have a concise presentation, the dependence of a system matrix of the subsystem ${\bf{\Sigma}}_i$ on its parameter matrix $P(i)$ is usually not explicitly expressed, except when this omission may cause some significant confusions.

\textcolor{black}{Substituting Equation (\ref{eqn:2}) into Equation (\ref{eqn:1}), the aforementioned NDS model takes completely the same form as the generalized state space model suggested in \cite{cw2020}, except that all the associated matrices are restricted to be block diagonal or the product of a block diagonal matrix and the SCM $\Phi$. This may mean that the model adopted in this paper is more flexible and more informative than that model of \cite{cw2020} in NDS structure descriptions. On the other hand, when the internal output vector $z(t,i)$ is equal to the external output vector $y(t,i)$ in each subsystem ${\bf{\Sigma}}_i$, $1\leq i\leq N$, and/or the SCM $\Phi$ is restricted to be symmetric, etc., this NDS model reduces to those adopted in \cite{cw2017,ptt2019,vtc2021}. }

To clarify dependence of the NDS ${\bf{\Sigma}}$, the external output vector of its $i$-th subsystem ${\bf{\Sigma}}_{i}$, etc., on the NDS SCM $\Phi$, they are sometimes also written as ${\bf{\Sigma}}(\Phi)$, $y(t,i,\Phi)$, etc.

Throughout this paper, the following assumptions are adopted.

\renewcommand{\labelenumi}{A.\arabic{enumi})}
\begin{enumerate}
\item The vectors $u(t,i)$, $v(t,i)$, $x(t,i)$, $y(t,i)$ and $z(t,i)$  respectively have a dimension of $m_{{\rm\bf u}i}$, $m_{{\rm\bf v}i}$, $m_{{\rm\bf x}i}$, $m_{{\rm\bf y}i}$ and $m_{{\rm\bf z}i}$.
\item Every NDS subsystem, that is, ${\bf{\Sigma}}_i$ with $i\in\{1,2,\cdots,N\}$, is well-posed, which is equivalent to that the matrix $I_{m_{{\rm\bf g}i}} - {G{(i)}}{P{(i)}}$ is invertible.
\item The NDS ${\bf{\Sigma}}$ itself is well-posed, which is equivalent to that the matrix $I-\Phi
{\rm\bf diag}\!\left\{\!A_{\rm\bf zv}(i)|_{i=1}^{N}\!\right\}$ is invertible.
\end{enumerate}

The first assumption is introduced to clarify vector size, while well-posedness of a system means that its states respond solely to each pair of their initial values and external inputs. That is, Assumptions A.2) and A.3) are necessary for a system to properly work  \cite{Kailath1980,sbkkmpr2011,zdg1996,zyl2018}. It can therefore be declared that all these three assumptions should be met by a practical system. In other words, the assumptions adopted here are not quite restrictive.

\begin{definition}
\textcolor{black}{The structure of the NDS of Equations (\ref{eqn:1}) and (\ref{eqn:2}) is identifiable}, if for an arbitrary initial state vector  ${\rm\bf col}\left\{\left. x(0,i)\right|_{i=1}^{i=N}\right\}$ and any two distinctive SCMs $\Phi_{1}$ and $\Phi_{2}$ satisfying Assumption A.3), there exists at least one external input time series $\left.{\rm\bf col}\left\{\left. u(t,i)\right|_{i=1}^{i=N}\right\}\right|_{t=0}^{t=\infty}$, such that the external output time series $\left.{\rm\bf col}\left\{\left. y(t,i,\Phi_{1})\right|_{i=1}^{i=N}\right\}\right|_{t=0}^{t=\infty}$ of the NDS ${\bf{\Sigma}}(\Phi_{1})$ is different from the external output time series $\left.{\rm\bf col}\left\{\left. y(t,i,\Phi_{2})\right|_{i=1}^{i=N}\right\}\right|_{t=0}^{t=\infty}$ of the NDS ${\bf{\Sigma}}(\Phi_{2})$. Otherwise, \textcolor{black}{the structure of this NDS is called unidentifiable.}
\label{def:1}
\end{definition}

From this definition, it is clear that if \textcolor{black}{the structure of an NDS is not identifiable}, then its subsystem interactions can not be determined through only experiments and the subsequent estimations, no matter what probing signals are used to stimulate the NDS, how long an experiment data length is, and what estimation algorithm is adopted. In other words, for this NDS, experiments only are not informative enough to distinguish its structure. This means that structure identifiability defined above is a property held by an NDS.

\textcolor{black}{The objectives of this paper are to develop a condition for verifying NDS structure identifiability that is both attractive from a computation respect and scalable with its subsystem number, as well as to derive requirements on each subsystem from which an NDS can be constructed whose structure can be estimated from experiment data.}

\textcolor{black}{Identifiability is an important issue in system identification, and has been attracting extensive attentions for a long time in various fields \cite{adm2020,wvd2018,zyl2018}. While various definitions have been given and many verification methods have been developed, this terminology traditionally and basically means parameter identifiability, and is sometimes also called a priori identifiability or structural identifiability. The former reflects the fact that identifiability is an essential prerequisite on a system model before its estimation, while the latter means that identifiability is generally a generic property of a system model (rather than identifiability of the structure of a system).  In general, identifiability verification is mathematically difficult and still remains challenging, even for an LTI system \cite{adm2020,vdhb2013,vtc2021,wvd2018}.}

\textcolor{black}{On the other hand, when a subsystem interaction is regarded as a system parameter, the above structure identifiability becomes the traditional one. As argued in \cite{cw2020,pmssaxcas2010,Siljak1978,sbkkmpr2011,vdhb2013,vtc2021,zyl2018} and the references therein, structure information is important in NDS analysis and synthesis. To reflect this aspect and be consistent with \cite{cw2020,Siljak1978,xz2014,zyl2018}, the terminology "structure identifiability" is adopted in this paper, which is also called topology identifiability in \cite{vtc2021} and network identifiability in \cite{vdhb2013,wvd2018}. In addition, note that in the analysis and synthesis of a large scale NDS, initial states of the NDS are usually not very clear, and the data length of an experiment is often not very large. These two factors have also been taken into account in the above definition.}

To develop a computationally feasible condition for verifying NDS structure identifiability, the following results are introduced \cite{Gantmacher1959,hj1991,zz2020}.

\begin{lemma}
Divide a matrix $A$ as $A = \left[ A_{1}^{T} \;  A_{2}^{T}\right]^{T}$, and assume that $A_{1}$ is not of FCR. Then the matrix $A$ is of full column rank (FCR), if and only if the matrix $A_{2}A_{1}^{\perp}$ is.
\label{lemma:1}
\end{lemma}

When the matrix $A_{1}$ is of FCR, $A_{1}^{\perp}=0$. In this case, for an arbitrary matrix $A_{2}$ with a compatible dimension, the matrix $A = \left[ A_{1}^{T} \;  A_{2}^{T}\right]^{T}$ is of FCR obviously.

\begin{lemma}
Assume that $A_{i}^{[j]}|_{i=1,j=1}^{i=3,j=m}$ and $B_{i}^{[j]}|_{i=1,j=1}^{i=3,j=m}$ are some matrices having compatible dimensions, and the matrix $\left[A_{2}^{[1]}\;A_{2}^{[2]}\;\cdots\; A_{2}^{[m]}\right]$ is of FCR. Then the matrix
\begin{displaymath}
\left[\!\!\begin{array}{ccc}
{\rm\bf diag}\!\left\{\!A_{1}^{[1]},\;A_{2}^{[1]},\; A_{3}^{[1]}\!\right\} &  \cdots &
{\rm\bf diag}\!\left\{\!A_{1}^{[m]},\;A_{2}^{[m]},\; A_{3}^{[m]}\!\right\}   \\
\left[ B_{1}^{[1]} \;\;\;\; B_{2}^{[1]} \;\;\;\; B_{3}^{[1]}\right] &  \cdots &
\left[ B_{1}^{[m]} \;\;\;\; B_{2}^{[m]} \;\;\;\; B_{3}^{[m]}\right]\end{array}\!\!\right]
\end{displaymath}
is of FCR, if and only if the following matrix has this property
\begin{displaymath}
\left[\begin{array}{ccc}
{\rm\bf diag}\left\{A_{1}^{[1]},\; A_{3}^{[1]}\right\} &  \cdots &
{\rm\bf diag}\left\{A_{1}^{[m]},\; A_{3}^{[m]}\right\}   \\
\left[ B_{1}^{[1]} \;\;\;\;  B_{3}^{[1]}\right] &  \cdots &
\left[ B_{1}^{[m]} \;\;\;\;  B_{3}^{[m]}\right]\end{array}\right]
\end{displaymath}
\label{lemma:2}
\end{lemma}

The following definitions and results are well known on matrix pencils, which can be found in many published works including \cite{bv1988,it2017}.

\begin{definition}
Let $G$ and $H$ be two arbitrary $m\times n$ dimensional real matrices. A  matrix valued polynomial (MVP) $\Psi(\lambda)=\lambda G+H$ is called a matrix pencil.
\begin{itemize}
\item This matrix pencil is called regular, whenever $m=n$ and \textcolor{black}{there exists a $\lambda\in {\cal C}$, such that ${\rm\bf det}(\Psi(\lambda))\neq 0$.}
\item If both the matrices $G$ and $H$ are invertible, then this matrix pencil is called strictly regular.
\item If there exist two nonsingular real matrices $U$ and $V$, such that  $\Psi(\lambda)=U\overline{\Psi}(\lambda)V$ is satisfied \textcolor{black}{at each  $\lambda\in {\cal C}$} by two matrix pencils ${\Psi}(\lambda)$ and $\overline{\Psi}(\lambda)$, then these two matrix pencils are said to be strictly equivalent.
\end{itemize}
\end{definition}

The following symbols are adopted throughout this paper. For an arbitrary positive integer $m$, the symbol $H_{m}(\lambda)$ stands for an $m\times m$ dimensional strictly regular matrix pencil, while the symbols $K_{m}(\lambda)$, $N_{m}(\lambda)$, $L_{m}(\lambda)$ and $J_{m}(\lambda)$ respectively for matrix pencils having the following definitions,
\begin{eqnarray}
& &
\hspace*{-1.0cm} K_{m}(\lambda)\!=\!\lambda I_{m}\!+\!\left[\!\!\begin{array}{cc}
0 & I_{m-1} \\ 0 & 0 \end{array}\!\!\!\right]\!,\hspace{0.1cm}
N_{m}(\lambda)\!=\!\lambda \!\left[\!\!\begin{array}{cc}
0 & I_{m-1} \\ 0 & 0 \end{array}\!\!\!\right] \!+\! I_{m} \label{eqn:3} \\
& &
\hspace*{-1.0cm} L_{m}(\lambda)=\left[\begin{array}{cc}
K_{m}(\lambda) & \left[\begin{array}{c} 0 \\ 1 \end{array}\right] \end{array}\right],\hspace{0.15cm}
J_{m}(\lambda)= \left[\begin{array}{c}
K_{m}^{T}(\lambda) \\ \left[0 \;\;\;\;\; 1\right] \end{array}\right] \label{eqn:4}
\end{eqnarray}
These matrix pencils are often used in constructing the Kronecker canonical form (KCF) of a general matrix pencil. Obviously, the dimensions of the matrix pencils $K_{m}(\lambda)$ and $N_{m}(\lambda)$ are $m\times m$, while the matrix pencils $L_{m}(\lambda)$ and $J_{m}(\lambda)$ respectively have a dimension of $m\times (m+1)$ and $(m+1)\times m$. Moreover, when $m=0$, $L_{m}(\lambda)$ is a $0\times 1$ zero matrix whose existence means adding a zero column vector in a KCF without increasing its rows, while $J_{m}(\lambda)$ is a $1\times 0$ zero matrix whose existence means adding a zero row vector in a KCF without increasing its columns. On the other hand, $J_{m}(\lambda)=L_{m}^{T}(\lambda)$. For a clear presentation, however, it appears better to introduce these two matrix pencils simultaneously.

In other words, throughout this paper, the capital letters $H$, $K$, $N$, $J$ and $L$ are used to indicate the type of the associated matrix pencil, while the subscript $m$ its dimensions.

When a matrix pencil is block diagonal with the diagonal blocks having the form $H_{*}(\lambda)$, $K_{*}(\lambda)$, $N_{*}(\lambda)$, $L_{*}(\lambda)$ and $J_{*}(\lambda)$, it is called KCF. It is now extensively known that any matrix pencil is strictly equivalent to a KCF \cite{bv1988,Gantmacher1959,it2017}. More precisely, we have the following results.

\begin{lemma}
For any matrix pencil $\Psi(\lambda)$, there are some unique nonnegative integers $\xi_{\rm\bf H}$, $\zeta_{\rm\bf K}$, $\zeta_{{\rm\bf L}}$, $\zeta_{{\rm\bf N}}$, $\zeta_{{\rm\bf J}}$, $\xi_{\rm\bf L}(j)|_{j=1}^{\zeta_{{\rm\bf L}}}$ and $\xi_{\rm\bf J}(j)|_{j=1}^{\zeta_{{\rm\bf J}}}$, as well as some unique positive integers $\xi_{\rm\bf K}(j)|_{j=1}^{\zeta_{{\rm\bf K}}}$ and $\xi_{\rm\bf N}(j)|_{j=1}^{\zeta_{{\rm\bf N}}}$, such that $\Psi(\lambda)$ is strictly equivalent to the block diagonal matrix pencil $\overline{\Psi}(\lambda)$ defined as
\begin{eqnarray}
\overline{\Psi}(\lambda)
\!\!\!\!&=&\!\!\!\!
{\rm\bf diag}\!\left\{\!H_{\xi_{{\rm\bf H}}}(\lambda),\;K_{\xi_{\rm\bf K}(j)}(\lambda)|_{j=1}^{\zeta_{{\rm\bf K}}},\; L_{\xi_{\rm\bf L}(j)}(\lambda)|_{j=1}^{\zeta_{{\rm\bf L}}}, \right.\nonumber\\
& & \hspace*{2cm}\left. N_{\xi_{\rm\bf N}(j)}(\lambda)|_{j=1}^{\zeta_{{\rm\bf N}}},\; J_{\xi_{\rm\bf J}(j)}(\lambda)\!|_{j=1}^{\zeta_{{\rm\bf J}}}\!\right\}
\label{eqn:5}
\end{eqnarray}
\label{lemma:3}
\end{lemma}

The following results are obtained in \cite{zz2020}, which explicitly characterizes the null spaces of the matrix pencils $H_{*}(\lambda)$, $K_{*}(\lambda)$, $N_{*}(\lambda)$, $L_{*}(\lambda)$ and $J_{*}(\lambda)$. This characterization is helpful in clarifying subsystems with which \textcolor{black}{an NDS can be constructed with its structure being identifiable.}

\begin{lemma} Let $m$ be an arbitrary positive integer. Then the matrix pencils defined respectively in Equations (\ref{eqn:3}) and (\ref{eqn:4}) have the following null spaces.
\begin{itemize}
\item $H_{m}(\lambda)$ is not of full rank (FR) only at $m$ isolated complex values of the variable $\lambda$. All these values are not equal to zero.
\item $N_{m}(\lambda)$ is always of FR.
\item $J_{m}(\lambda)$ is always of FCR.
\item $K_{m}(\lambda)$ is singular only at $\lambda=0$, and $K_{m}^{\perp}(0)={\rm\bf col}\left\{1, 0_{m-1}\right\}$.
\item $L_{m}(\lambda)$ is not of FCR at any complex $\lambda$, and
$L_{m}^{\perp}(\lambda)={\rm\bf col}\left\{\left. 1, (-\lambda)^{j}\right|_{j=1}^{m}\right\}$.
\end{itemize}
\label{lemma:4}
\end{lemma}

\section{NDS Structure Identifiability}

To establish conditions on a subsystem such that \textcolor{black}{the structure of an NDS constituted from it is identifiable}, for each subsystem ${\rm\bf\Sigma}_{i}$ of the NDS ${\rm\bf\Sigma}$, in which $i=1,2,\cdots,N$,
define TFMs $G_{\rm\bf zu}(\lambda,i)$, $G_{\rm\bf zv}(\lambda,i)$, $G_{\rm\bf yu}(\lambda,i)$ and $G_{\rm\bf yv}(\lambda,i)$ respectively as
\begin{eqnarray*}
& & \hspace*{-0.8cm} \left[\!\!\begin{array}{cc}
G_{\rm\bf yu}(\lambda,i) & G_{\rm\bf yv}(\lambda,i)  \\
G_{\rm\bf zu}(\lambda,i) & G_{\rm\bf zv}(\lambda,i)
\end{array}\!\!\right]
\!\!=\!\!
\left[\!\!\begin{array}{cc}
D_{\rm\bf u}(i) & C_{\rm\bf v}(i) \\
B_{\rm\bf z}(i) & A_{\rm\bf zv}(i)
\end{array}\!\!\right] +
\left[\!\!\begin{array}{c}
C_{\rm\bf x}(i) \\
A_{\rm\bf zx}(i)
\end{array}\!\!\right]\!\times \\
& & \hspace*{2.5cm} \left[\lambda I_{m_{{\rm\bf x}i}}-A_{\rm\bf xx}(i)\right]^{-1}
\left[\!\!\begin{array}{cc}
B_{\rm\bf x}(i) & A_{\rm\bf xv}(i)
\end{array}\!\!\right]
\end{eqnarray*}
in which $\lambda$ stands for the Laplace transformation variable $s$ when the NDS $\rm\bf\Sigma$ is of continuous time, and for the ${\cal Z}$ transformation variable $z$ when the NDS $\rm\bf\Sigma$ is of discrete time. Moreover, define block diagonal TFMs $G_{\rm\bf \star\#}(\lambda)$ with ${\rm\bf\star}={\rm\bf z}$ or ${\rm\bf y}$ and ${\rm\bf\#}={\rm\bf u}$ or ${\rm\bf v}$ as
\begin{displaymath}
G_{\star\#}(\lambda) = {\rm\bf diag}\!\left\{G_{\rm\bf
\star\#}(\lambda,i)|_{i=1}^{N}\!\right\}
\end{displaymath}

Note that the well-posedness of the NDS $\rm\bf\Sigma$ is equivalent to that the matrix $I_{m_{\rm\bf z}} - A_{\rm\bf zv}\Phi$ is invertible, in which $A_{\rm\bf zv}={\rm\bf diag}\!\left\{A_{\rm\bf zv}(i)|_{i=1}^{N}\!\right\}$ and $ m_{\rm\bf z}={\sum_{k=1}^{N} m_{{\rm\bf z}k}}$ \cite{Zhou2015,zz2020}. On the other hand, define matrices $A_{\rm\bf zx}$, $A_{\rm\bf xx}$ and $A_{\rm\bf xv}$ respectively as $A_{\rm\bf zx}={\rm\bf diag}\!\left\{A_{\rm\bf zx}(i)|_{i=1}^{N}\!\right\}$,
$A_{\rm\bf xx}={\rm\bf diag}\!\left\{A_{\rm\bf xx}(i)|_{i=1}^{N}\!\right\}$ and
$A_{\rm\bf xv}={\rm\bf diag}\!\left\{A_{\rm\bf xv}(i)|_{i=1}^{N}\!\right\}$. Moreover, denote ${\sum_{k=1}^{N} m_{{\rm\bf x}k}}$ by $m_{\rm\bf x}$. Then when the NDS $\rm\bf\Sigma$ satisfies Assumption A.3), from the block diagonal structure of the TFM $G_{\rm\bf zv}(\lambda)$, we have that
\begin{eqnarray}
& &\!\!\!\! I_{m_{\rm\bf z}} - G_{\rm\bf zv}(\lambda)\Phi  \nonumber\\
&=&\!\!\!\! I_{m_{\rm\bf z}} - \left\{A_{\rm\bf zv} + A_{\rm\bf zx}\left[\lambda I_{m_{\rm\bf x}}-A_{\rm\bf xx}\right]^{-1}A_{\rm\bf xv}\right\} \Phi \nonumber\\
&=&\!\!\!\! \left(I_{m_{\rm\bf z}}-A_{\rm\bf zv}\Phi\right)\!\!\left\{\!I_{m_{\rm\bf z}} \!-\!
\left(I_{m_{\rm\bf z}} \!-\! A_{\rm\bf zv}\Phi\right)^{-1}\!\!\times \right.\nonumber\\
& &\hspace*{3cm} \left. A_{\rm\bf zx}\left[\lambda I_{m_{\rm\bf x}} \!-\! A_{\rm\bf xx}\right]^{-1}\!\!A_{\rm\bf xv}\Phi \! \right\}
\label{eqn:15}
\end{eqnarray}
Hence, from the determinant equality ${\rm\bf det}(I-AB)={\rm\bf det}(I-BA)$ which is well known in matrix theories \cite{Gantmacher1959,hj1991}, we have that
\begin{eqnarray}
& &\!\!\!\! {\rm\bf det}\!\left\{I_{m_{\rm\bf z}} - G_{\rm\bf zv}(\lambda)\Phi \right\}  \nonumber\\
&=&\!\!\!\! {\rm\bf det}\!\left(I_{m_{\rm\bf z}}-A_{\rm\bf zv}\Phi\right)\times \nonumber\\
& &\!\!\!\! {\rm\bf det}\!\left\{\!\!I_{m_{\rm\bf z}} \!-\!
\left(I_{m_{\rm\bf z}} \!-\! A_{\rm\bf zv}\Phi\right)^{-1} A_{\rm\bf zx}\left[\lambda I_{m_{\rm\bf x}} \!-\! A_{\rm\bf xx}\right]^{-1}\!\!A_{\rm\bf xv}\Phi\!\! \right\} \nonumber\\
&=&\!\!\!\! {\rm\bf det}\!\left(I_{m_{\rm\bf z}}-A_{\rm\bf zv}\Phi\right)\times \nonumber\\
& &\!\!\!\! {\rm\bf det}\!\left\{\!\!I_{m_{\rm\bf x}} \!-\!
A_{\rm\bf xv}\Phi\left(I_{m_{\rm\bf x}} \!-\! A_{\rm\bf zv}\Phi\right)^{-1} A_{\rm\bf zx}\left[\lambda I_{m_{\rm\bf x}} \!-\! A_{\rm\bf xx}\right]^{-1}\!\! \right\} \nonumber\\
&=&\!\!\!\!{\rm\bf det}^{-1}\!\left(\lambda I_{m_{\rm\bf x}} \!-\! A_{\rm\bf xx}\right) \times {\rm\bf det}\!\left(I_{m_{\rm\bf z}}-A_{\rm\bf zv}\Phi\right)\times \nonumber\\
& &\!\!\!\! {\rm\bf det}\!\left\{\!\lambda I_{m_{\rm\bf x}} \!-\! \left[A_{\rm\bf xx} \!+\!
A_{\rm\bf xv}\Phi\left(I_{m_{\rm\bf x}} \!-\! A_{\rm\bf zv}\Phi\right)^{-1} A_{\rm\bf zx}\!\right]\! \right\}
\label{eqn:16}
\end{eqnarray}

Recall that all the matrices and TFMs in the above equation are of a finite dimension. This means that when the NDS $\rm\bf\Sigma$ is well-posed, ${\rm\bf det}\!\left\{I_{m_{\rm\bf z}} - G_{\rm\bf zv}(\lambda)\Phi \right\}$ is not \textcolor{black}{identically} equal to zero. That is, the TFM $I_{m_{\rm\bf z}} - G_{\rm\bf zv}(\lambda)\Phi$ is of full normal rank (FNR). Hence, its inverse is well-defined. On the basis of these observations, define a SCM $\Phi$ dependent TFM $H(\lambda,\Phi)$ as
\begin{equation}
\hspace*{-0.0cm} H(\lambda,\Phi) \!=\! G_{\rm\bf yu}(\lambda) + G_{\rm\bf yv}(\lambda)\Phi\left[I_{m_{\rm\bf z}} \!-\! G_{\rm\bf zv}(\lambda)\Phi\right]^{-1}\!\!G_{\rm\bf zu}(\lambda)
\label{eqn:6}
\end{equation}
Then the following results can be established for the structure identifiability of the NDS $\rm\bf\Sigma$, while their proof is deferred to the appendix.

\begin{theorem}
Assume that the NDS $\rm\bf\Sigma$ is well-posed. \textcolor{black}{Then its structure is identifiable}, if and only if for each SCM pair $\Phi_{1}$ and $\Phi_{2}$ satisfying $\Phi_{1} \neq \Phi_{2}$, $H(\lambda,\Phi_{2}) \neq H(\lambda,\Phi_{1})$ at almost every $\lambda\in {\cal C}$.
\label{theorem:1}
\end{theorem}

This theorem makes it clear that the structure identifiability studied in this paper is equivalent to that investigated in \cite{cw2017,cw2020,vtc2021}, in which \textcolor{black}{the structure of an NDS is called identifiable} if any two different SCMs lead to different TFMs of the whole system.

The necessity and sufficiency of the above condition are to some extent clear from an application viewpoint. Particularly, when there are two sets of subsystem interactions that lead to the same external outputs for each external stimulus, it is not out of imagination that these two subsystem interaction sets result in the same TFM of the whole NDS from its external inputs to its external outputs. On the other hand, if two distinctive SCMs lead to the same NDS TFM, the external outputs of the corresponding NDSs are usually hard to be distinguished when they are stimulated by the same external inputs.

On the basis of these results, as well as properties of an LFT, a computationally feasible condition is derived for the structure identifiability of the NDS $\rm\bf\Sigma$.

\begin{theorem}
Assume that the NDS $\rm\bf\Sigma$ satisfies Assumptions A.1)-A.3). If for each $i=1,2,\cdots,N$, the TFM $G_{\rm\bf yv}(\lambda,i)$ is of FNCR, while the TFM $G_{\rm\bf zu}(\lambda,i)$ is of FNRR, \textcolor{black}{then the structure of this NDS is identifiable.}
\label{theorem:2}
\end{theorem}

The proof of the above theorem is given in the appendix.

From this theorem, it is clear that structure identifiability of an NDS can be completely determined by the dynamics of its individual subsystems. This is quite attractive in NDS constructions including subsystem dynamics selection, external input/output position determination, etc., as well as experiment designs for NDS identification.

Theorem \ref{theorem:2} also makes it clear that through adding external outputs and/or external inputs, it is possible to change \textcolor{black}{the structure of an NDS from being unidentifiable to being identifiable}. Noting that adding an external input/output usually increases information in experiment data about an NDS structure, this observation is consistent well with intuitions in actual applications.

\textcolor{black}{Note that the conditions of Theorem \ref{theorem:2} depend only on the TFMs $G_{\rm\bf yv}(\lambda,i)$ and $G_{\rm\bf zu}(\lambda,i)$. It is straightforward to extend them to an NDS whose subsystem dynamics are described in a descriptor form. Compared with the results of \cite{ptt2019}, the conditions are only sufficient. On the other hand, there are less restrictions on subsystem dimensions and interactions, and the conditions can be verified with each subsystem independently. More specifically, the condition of \cite{ptt2019} is both necessary and sufficient. Each subsystem, however, is required to be square and subsystems are asked to be connected through their external outputs by a positive semidefinite SCM. In addition, the condition there depends on the SCM $\Phi$ which may in general not be very competitive in large scale NDS analysis and synthesis.}

When subsystem parameters are known, the associated subsystem matrices are completely determined and therefore the corresponding TFMs $G_{\rm\bf yv}(\lambda,i)|_{i=1}^{N}$ and $G_{\rm\bf zu}(\lambda,i)|_{i=1}^{N}$. Under such a situation, the condition of Theorem \ref{theorem:2} can be simply verified through directly investigating their Smith-McMillan forms, etc.

More precisely, let $r_{\rm\bf yv}^{[i]}$ and $r_{\rm\bf zu}^{[i]}$ with $i=1,2,\cdots,N$, stand respectively for the maximum ranks of the TFMs $G_{\rm\bf yv}(\lambda,i)$ and $G_{\rm\bf zu}(\lambda,i)$ when $\lambda$ varies over the set $\cal C$. Then it is obvious from the dimensions of these TFMs that $0\leq r_{\rm\bf yv}^{[i]}\leq \max\{m_{{\rm\bf v}i},\;m_{{\rm\bf y}i}\}$ and $0\leq r_{\rm\bf zu}^{[i]}\leq \max\{m_{{\rm\bf z}i},\;m_{{\rm\bf u}i}\}$. Moreover, their Smith-McMillan forms can be respectively written as follow,
\begin{equation}
G_{\rm\bf yv}(\lambda,i) \!=\! U_{\rm\bf yv}(\lambda,i)
\!\left[\!\!\!\begin{array}{cc}
{\rm\bf diag}\!\left\{\!\!\left.\frac{\alpha^{[j]}_{\rm\bf yv}(\lambda,i)}{\beta^{[j]}_{\rm\bf yv}(\lambda,i)}\!\right|_{j=1}^{r_{\rm\bf yv}^{[i]}}\!\!\right\} & 0 \\
0 & 0 \end{array}\!\!\!\right]\!
V_{\rm\bf yv}(\lambda,i)
\label{eqn:29}
\end{equation}
and
\begin{equation}
G_{\rm\bf zu}(\lambda,i) \!=\! U_{\rm\bf zu}(\lambda,i)
\!\!\left[\!\!\!\begin{array}{cc}
{\rm\bf diag}\!\left\{\!\!\left.\frac{\alpha^{[j]}_{\rm\bf zu}(\lambda,i)}{\beta^{[j]}_{\rm\bf zu}(\lambda,i)}\!\right|_{j=1}^{r_{\rm\bf zu}^{[i]}}\!\right\} & 0 \\
0 & 0 \end{array}\!\!\right]\!
V_{\rm\bf zu}(\lambda,i)
\label{eqn:30}
\end{equation}
in which the zero matrices in general have different dimensions, while
$U_{\rm\bf yv}(\lambda,i)$, $U_{\rm\bf zu}(\lambda,i)$, $V_{\rm\bf yv}(\lambda,i)$ and $V_{\rm\bf zu}(\lambda,i)$ are respectively $m_{{\rm\bf y}i}\times m_{{\rm\bf y}i}$, $m_{{\rm\bf z}i}\times m_{{\rm\bf z}i}$, $m_{{\rm\bf v}i}\times m_{{\rm\bf v}i}$ and $m_{{\rm\bf u}i}\times m_{{\rm\bf u}i}$ dimensional unimodular matrices, and   $\alpha^{[j]}_{\rm\bf yv}(\lambda,i)|_{j=1}^{r_{\rm\bf yv}^{[i]}}$, $\alpha^{[j]}_{\rm\bf zu}(\lambda,i)|_{j=1}^{r_{\rm\bf zu}^{[i]}}$,   $\beta^{[j]}_{\rm\bf yv}(\lambda,i)|_{j=1}^{r_{\rm\bf yv}^{[i]}}$ and $\beta^{[j]}_{\rm\bf zu}(\lambda,i)|_{j=1}^{r_{\rm\bf zu}^{[i]}}$ are real coefficient polynomials that are not \textcolor{black}{identically} equal to zero and have a finite degree.

As argued in the proof of Theorem \ref{theorem:2}, the TFM $G_{\rm\bf yv}(\lambda,i)$ is of FNCR, if and only if $r_{\rm\bf yv}^{[i]} = m_{{\rm\bf v}i}$. On the other hand, the TFM $G_{\rm\bf zu}(\lambda,i)$ is of FNRR, if and only if $r_{\rm\bf zu}^{[i]} = m_{{\rm\bf z}i}$. Note that the dimensions of a subsystem in an NDS are usually not very large. This means that the Smith-McMillan forms of the aforementioned TFMs can be easily obtained in general, and therefore the condition of Theorem \ref{theorem:2} can be checked without any significant difficulties.

While the above theorem gives a condition for the NDS structure identifiability which can be easily verified, it is only sufficient \textcolor{black}{and is in general not necessary which is illustrated by the numerical example of Section V.} In addition, this condition may not be satisfied easily in some applications, as it requires that in each NDS subsystem, the number of its external outputs is not smaller than that of its internal inputs, while the number of its external inputs is not smaller than that of its internal outputs. To settle these issues, the following results are derived, which give a necessary and sufficient condition for the structure identifiability of the NDS $\rm\bf\Sigma$, without any restrictions on the TFMs ${G}_{\rm\bf zu}(\lambda)$, ${G}_{\rm\bf yv}(\lambda)$ and ${G}_{\rm\bf zv}(\lambda)$.

\begin{corollary}
Let $N_{yv,r}(\lambda)D^{-1}_{yv,r}(\lambda)$ and $D^{-1}_{zu,l}(\lambda)N_{zu,l}(\lambda)$ respectively be a right coprime factorization of the TFM ${G}_{\rm\bf yv}(\lambda)$ and a left coprime factorization of the TFM ${G}_{\rm\bf zu}(\lambda)$. Assume that each subsystem of the NDS $\rm\bf\Sigma$, as well as the NDS itself, is well-posed. Denote the coefficient matrices of the MVP $N^{T}_{yv,r}(\lambda)\otimes N_{zu,l}(\lambda)$ by $\Xi_{\rm\bf zu}^{\rm\bf yv}(i)$ with $0\leq i \leq d_{\rm\bf zu}^{\rm\bf yv}$, in which $d_{\rm\bf zu}^{\rm\bf yv}$ stands for the degree of the aforementioned MVP. Then \textcolor{black}{the structure of the NDS $\rm\bf\Sigma$ is identifiable}, if and only if the matrix ${\rm\bf col}\!\left\{ \left.\Xi_{\rm\bf zu}^{\rm\bf yv}(i)\right|_{0}^{d_{\rm\bf zu}^{\rm\bf yv}}\right\}$ is of FCR.
\label{cor:0}
\end{corollary}

\hspace*{-0.45cm}{\rm\bf Proof:} Let $\Phi_{1}$ and $\Phi_{2}$ be two arbitrary SCMs satisfying the well-posedness assumption. From Equation (\ref{eqn:a7}), we immediately have that 
\begin{eqnarray}
& & \!\!\!\! {\rm\bf vec}\left\{H(\lambda,\Phi_{1}) - H(\lambda,\Phi_{2})\right\} \nonumber \\
&=& \!\!\!\! \left\{ \left[ G_{\rm\bf yv}(\lambda)\left[I_{m_{\rm\bf v}} \!-\! \Phi_{2}G_{\rm\bf zv}(\lambda)\right]^{-1} \right]^{T} \otimes \right. \nonumber\\
& & \hspace*{0.5cm} \left.\left[\left[I_{m_{\rm\bf z}} \!-\! G_{\rm\bf zv}(\lambda)\Phi_{1}\right]^{-1}\!\!G_{\rm\bf zu}(\lambda)\right]\right\} \!{\rm\bf vec}\left\{\Phi_{1} \!-\!\Phi_{2}\right\}  \nonumber\\
&=& \!\!\!\! \left\{ \left[ I_{m_{\rm\bf v}} \!-\! \Phi_{2}G_{\rm\bf zv}(\lambda)\right]^{T} \otimes 
\left[I_{m_{\rm\bf z}} \!-\! G_{\rm\bf zv}(\lambda)\Phi_{1}\right] \right\}^{-1} \times \nonumber\\
& & \hspace*{2.5cm}
\left\{G^{T}_{\rm\bf yv}(\lambda)\otimes G_{\rm\bf zu}(\lambda)\right\} \!{\rm\bf vec}\left\{\Phi_{1} \!-\!\Phi_{2}\right\}  \nonumber\\
&=& \!\!\!\! \left\{ \left[ I_{m_{\rm\bf v}} \!-\! \Phi_{2}G_{\rm\bf zv}(\lambda)\right]^{T} \otimes
\left[I_{m_{\rm\bf z}} \!-\! G_{\rm\bf zv}(\lambda)\Phi_{1}\right] \right\}^{-1} \times \nonumber\\
& & \hspace*{1.5cm}
\left\{D^{T}_{\rm\bf yv,r}(\lambda)\otimes D_{\rm\bf zu,l}(\lambda)\right\}^{-1} \times \nonumber\\
& & \hspace*{1.5cm}\left\{N^{T}_{\rm\bf yv,r}(\lambda)\otimes N_{\rm\bf zu,l}(\lambda)\right\} \!{\rm\bf vec}\left\{\Phi_{1} \!-\!\Phi_{2}\right\}
\end{eqnarray}

On the basis of the well-posedness assumption, we have that both $I_{m_{\rm\bf v}} \!-\! \Phi_{2}G_{\rm\bf zv}(\lambda)$ and $I_{m_{\rm\bf z}} \!-\! G_{\rm\bf zv}(\lambda)\Phi_{1}$ are invertible. It can therefore be declared that ${\rm\bf vec}\left\{H(\lambda,\Phi_{1}) - H(\lambda,\Phi_{2})\right\} \equiv 0$, which is equivalent to $H(\lambda,\Phi_{1}) \equiv H(\lambda,\Phi_{2})$, if and only if 
\begin{equation}
\left\{N^{T}_{\rm\bf yv,r}(\lambda)\otimes N_{\rm\bf zu,l}(\lambda)\right\} \!{\rm\bf vec}\left\{\Phi_{1} \!-\!\Phi_{2}\right\} = 0
\end{equation}
which can be further expressed as that for each $i=0,1,\cdots,d_{\rm\bf zu}^{\rm\bf yv}$, 
\begin{equation}
\Xi_{\rm\bf zu}^{\rm\bf yv}(i) \!{\rm\bf vec}\left\{\Phi_{1} \!-\!\Phi_{2}\right\} = 0
\end{equation}

The proof can now be completed noting that $\Phi_{1} \!=\!\Phi_{2}$, if and only if ${\rm\bf vec}\left\{\Phi_{1} \!-\!\Phi_{2}\right\} = 0$.   \hspace{\fill}$\Diamond$

The above corollary shows that structure identifiability of the NDS $\rm\bf\Sigma$ does not depend on a particular SCM $\Phi$, which is described by Equations (\ref{eqn:1}) and (\ref{eqn:2}). This means that the associated conditions are global, and are therefore attractive in NDS analysis and synthesis. In some actual NDSs, there may exist some constraints on the SCM $\Phi$. For example,
\begin{displaymath}
	\Phi(\theta)=\Phi^{[0]}+\sum^{q}_{k=1}\theta^{[k]}\Phi^{[k]}
\end{displaymath}
in which $\theta$ denotes the vector $col\{\theta^{[k]}|_{k=1}^{q}\}$, $\Phi^{[0]} \in \mathcal{R}^{m_v\times m_z}$ is a fixed constant matrix representing some a priori information on NDS structure, while $\Phi^{[k]} \in \mathcal{R}^{m_v\times m_z}$ with $k=1,2,\cdots,q$, is also a fixed constant matrix reflecting some known structure information about algebraic dependence among subsystem interactions. On the other hand, $\theta^{[k]}\in\mathbb{R}$ with $k=1,2,\cdots,q,$ is an unknown parameter to be estimated from experiment data, etc., and $q$ is the number of the algebraically independent parameters in the SCM $\Phi$. Conclusions of Corollary \ref{cor:0} can be easily extended to these situations.

For some particular kinds of NDSs, more simple conditions can be obtained for their structure identifiability.

For this purpose, partition the unimodular matrices $V_{\rm\bf yv}(\lambda,i)$ and $U_{\rm\bf zu}(\lambda,i)$ of Equations (\ref{eqn:29}) and (\ref{eqn:30}) respectively as
\begin{displaymath}
V_{\rm\bf yv}(\lambda,i) \!=\!\!
\left[\!\!\!\begin{array}{c}
V_{\rm\bf yv}^{[1]}(\lambda,i) \\
V_{\rm\bf yv}^{[2]}(\lambda,i) \end{array}\!\!\!\right]\!, \hspace{0.25cm}
U_{\rm\bf zu}(\lambda,i) \!=\!\!
\left[\!\!\!\begin{array}{cc}
U_{\rm\bf zu}^{[1]}(\lambda,i) &
U_{\rm\bf zu}^{[2]}(\lambda,i) \end{array}\!\!\!\right]
\end{displaymath}
in which the sub-MVP $V_{\rm\bf yv}^{[1]}(\lambda,i)$ has $r_{\rm\bf yv}^{[i]}$ rows, the sub-MVP $U_{\rm\bf zu}^{[1]}(\lambda,i)$ has $r_{\rm\bf zu}^{[i]}$ columns, while the other sub-MVPs have compatible dimensions. Moreover, denote the highest degrees of the MVP $V_{\rm\bf yv}^{[1]}(\lambda,i)$ and the MVP $U_{\rm\bf zu}^{[1]}(\lambda,i)$ respectively by $d^{[vi1]}_{\rm\bf yv}$ and $d^{[ui1]}_{\rm\bf zu}$. Then these two MVPs can be rewritten as
\begin{equation}
V_{\rm\bf yv}^{[1]}(\lambda,i)\!=\!\!\sum_{j=0}^{d^{[vi1]}_{\rm\bf yv}}
V_{\rm\bf yv}^{[1]}(i,j)\lambda^{j} , \hspace{0.25cm}
U_{\rm\bf zu}^{[1]}(\lambda,i) \!=\!\!
\sum_{j=0}^{d^{[ui1]}_{\rm\bf zu}}
U_{\rm\bf zu}^{[1]}(i,j)\lambda^{j}
\label{eqn:31}
\end{equation}
in which $V_{\rm\bf yv}^{[1]}(i,j)|_{j=1}^{d^{[vi1]}_{\rm\bf yv}}$ and $U_{\rm\bf zu}^{[1]}(i,j)|_{j=1}^{d^{[ui1]}_{\rm\bf zu}}$ are respectively $r_{\rm\bf yv}^{[i]}\times m_{{\rm\bf v}i}$ dimensional and $m_{{\rm\bf z}i}\times r_{\rm\bf zu}^{[i]}$ dimensional real matrices. Furthermore, for each $i,j=1,2,\cdots,N$, denote the following matrix by $\Xi_{\rm\bf zu}^{\rm\bf yv}(i,j)$,
\begin{displaymath}
{\rm\bf col}\!\left\{\!\!
\left. {\displaystyle
\sum_{s=\max\{0,k-d^{[uj1]}_{\rm\bf zu}\}}^{\min\{k,d^{[uj1]}_{\rm\bf zu}\}}
\!\!\!\!\!\!\!\!\!\! U_{\rm\bf zu}^{[1]T}(i,k \!-\!s) \otimes V_{\rm\bf yv}^{[1]}(j,s)} \right|_{k=0}^{d^{[uj1]}_{\rm\bf zu}+d^{[vi1]}_{\rm\bf yv}} \!\right\}
\end{displaymath}

With these symbols, the following results are obtained for the structure identifiability of the NDS of Equations (\ref{eqn:1}) and (\ref{eqn:2}), while their proof is postponed to the appendix.

\begin{theorem}
Assume that each subsystem of the NDS $\rm\bf\Sigma$ is well-posed. Moreover, assume that its subsystem TFM $G_{\rm\bf zv}(\lambda,i)$ is \textcolor{black}{identically} equal to zero for each $i=1,2,\cdots,N$. Then \textcolor{black}{the structure of this NDS is identifiable}, if and only if for every $i,j=1,2,\cdots,N$, the matrix $\Xi_{\rm\bf zu}^{\rm\bf yv}(i,j)$ is of FCR.
\label{theorem:5}
\end{theorem}

Note that the condition of this theorem can be checked for \textcolor{black}{each tuple of subsystems independently, its computational complexity increases only quadratically with the subsystem number, and is therefore scalable for an NDS with its subsystem number}. This is attractive in large scale NDS analysis and synthesis, in which the scalability of a condition is essential from computational considerations. On the other hand, the proof of the above theorem also reveals that in addition to using the Smith-McMillan forms of the TFMs $G_{\rm\bf yv}(\lambda,i)$ and $G_{\rm\bf zu}(\lambda,i)$, similar results can be obtained through their left and right coprime matrix polynomial descriptions, which have also been widely adopted as a plant model in system analysis and synthesis \cite{Kailath1980,zdg1996}. The details are omitted due to their obviousness.

\textcolor{black}{While the condition is quite restrictive and may not be easily satisfied in actual applications that TFM $G_{\rm\bf zv}(\lambda,i)\equiv 0$ for each $1\leq i\leq N$, this condition does not mean that in the NDS $\rm\bf\Sigma$, each subsystem is isolated from other subsystems. This can be seen from its TFM given by Equation (\ref{eqn:6}). Moreover, the derivations of Theorem \ref{theorem:5} are quite helpful in getting conditions for structure identifiability of another class of NDSs which includes those of \cite{cw2017,cw2020,ptt2019,vtc2021} as a special case. More precisely, in addition to the situation of Theorem \ref{theorem:5}}, a similar necessary and sufficient condition can also be established using similar arguments for NDS structure identifiability, provided that the TFM from the internal inputs to the external outputs of each subsystem can be expressed as a FNCR TFM multiplied by the TFM from its internal inputs to its internal outputs, while the TFM from the external inputs to the internal outputs of each subsystem can be expressed as that TFM multiplied by a FNRR TFM.

In particular, for each $i,j\in\{1,2,\cdots,N\}$, define a matrix $\Xi_{\rm\bf zv}^{\rm\bf zv}(i,j)$ as its counterpart $\Xi_{\rm\bf zu}^{\rm\bf yv}(i,j)$, using the Smith-McMillan form of the TFM $G_{\rm\bf zv}(\lambda,i)$ and that of the TFM $G_{\rm\bf zv}(\lambda,j)$. Then through similar arguments as those in the proof of Theorem \ref{theorem:5}, a necessary and sufficient condition is obtained under the aforementioned situation for the structure identifiability of the NDS $\rm\bf\Sigma$ of Equations (\ref{eqn:1}) and (\ref{eqn:2}).

\begin{corollary}
Assume that each subsystem of the NDS $\rm\bf\Sigma$, as well as the NDS itself, is well-posed. Moreover, assume that there exist a FNCR TFM $\overline{G}_{\rm\bf yv}(\lambda)$ and a FNRR TFM $\overline{G}_{\rm\bf zu}(\lambda)$, such that
\begin{equation}
G_{\rm\bf yv}\!(\lambda) \!=\! \overline{G}_{\rm\bf yv}\!(\lambda)G_{\rm\bf zv}(\lambda),\hspace{0.15cm} G_{\rm\bf zu}\!(\lambda) \!=\! G_{\rm\bf zv}\!(\lambda)\overline{G}_{\rm\bf zu}\!(\lambda)
\label{eqn:34}
\end{equation}
Then \textcolor{black}{the structure of this NDS is identifiable}, if and only if for every $i,j=1,2,\cdots,N$, the matrix $\Xi_{\rm\bf zv}^{\rm\bf zv}(i,j)$ is of FCR.
\label{cor:2}
\end{corollary}

\hspace*{-0.45cm}{\rm\bf Proof:} Let $\Phi_{1}$ and $\Phi_{2}$ be two arbitrary SCMs satisfying the well-posedness assumption. Substitute Equation (\ref{eqn:34}) into Equation (\ref{eqn:a7}). Then from Equation (\ref{eqn:6}), we have that
\begin{eqnarray}
& & \!\!\!\! H(\lambda,\Phi_{1}) - H(\lambda,\Phi_{2}) \nonumber \\
&=& \!\!\!\!  \overline{G}_{\rm\bf yv}(\lambda)G_{\rm\bf zv}(\lambda)\left[I_{m_{\rm\bf v}} \!-\! \Phi_{2}G_{\rm\bf zv}(\lambda)\right]^{-1} \!\left(\Phi_{1} \!-\!\Phi_{2}\right) \times \nonumber\\
& & \hspace*{1.5cm} \left[I_{m_{\rm\bf z}} \!-\! G_{\rm\bf zv}(\lambda)\Phi_{1}\right]^{-1}\!\!G_{\rm\bf zv}(\lambda)\overline{G}_{\rm\bf zu}(\lambda)  \nonumber\\
&=& \!\!\!\!  \overline{G}_{\rm\bf yv}(\lambda)\left[I_{m_{\rm\bf z}} \!-\! G_{\rm\bf zv}(\lambda)\Phi_{2}\right]^{-1} \!G_{\rm\bf zv}(\lambda)\left(\Phi_{1} \!-\!\Phi_{2}\right)\!G_{\rm\bf zv}(\lambda)\! \times \nonumber\\
& & \hspace*{1.5cm} \left[I_{m_{\rm\bf v}} \!-\! \Phi_{1}G_{\rm\bf zv}(\lambda)\right]^{-1}\!\!\overline{G}_{\rm\bf zu}(\lambda)
\label{eqn:35}
\end{eqnarray}

Now assume that the TFMs $\overline{G}_{\rm\bf yv}(\lambda)$ and $\overline{G}_{\rm\bf zu}(\lambda)$ are respectively of FNCR and FNRR. Noting that both the TFMs $\left[I_{m_{\rm\bf v}} \!-\! \Phi_{1}G_{\rm\bf zv}(\lambda)\right]^{-1}$ and $\left[I_{m_{\rm\bf z}} \!-\! G_{\rm\bf zv}(\lambda)\Phi_{2}\right]^{-1}$ are of FNR, it can be straightforwardly declared that
the TFMs $\overline{G}_{\rm\bf yv}(\lambda)\left[I_{m_{\rm\bf z}} \!-\! G_{\rm\bf zv}(\lambda)\Phi_{2}\right]^{-1}$ and $\left[I_{m_{\rm\bf v}} \!-\! \Phi_{1}G_{\rm\bf zv}(\lambda)\right]^{-1}\overline{G}_{\rm\bf zu}(\lambda)$ are also respectively of FNCR and FNRR.

From these observations, it can be proven using similar arguments as those in the proof of Theorem \ref{theorem:2} that $H(\lambda,\Phi_{1}) - H(\lambda,\Phi_{2})\equiv 0$, if and only if $G_{\rm\bf zv}(\lambda)\left(\Phi_{1} \!-\!\Phi_{2}\right)\!G_{\rm\bf zv}(\lambda)\equiv 0$.

The proof can now be completed by the same token as those in proving Theorem \ref{theorem:5}.   \hspace{\fill}$\Diamond$

Note that the matrix $\Xi_{\rm\bf zv}^{\rm\bf zv}(i,j)$ of Corollary \ref{cor:2} shares the same structure with the matrix $\Xi_{\rm\bf zu}^{\rm\bf yv}(i,j)$ of Theorem \ref{theorem:5}. It is obvious that the condition of Corollary \ref{cor:2} has the same computational advantages as that of Theorem \ref{theorem:5}. That is, its computational costs increase in general quadratically with the NDS subsystem number.

In some real world problems, there may have some {\it a priori} structure information about subsystem connections. For example, some elements of the SCM $\Phi$ must be fixed to be zero, which means that some subsystem internal outputs are prohibited to directly affect some subsystem internal inputs, etc. The conditions of Theorem \ref{theorem:5} and Corollary \ref{cor:2} can be easily modified to these situations. In particular, it is straightforward from Equation (\ref{eqn:a46}) that under the aforementioned condition, the only required modifications are to remove the corresponding column(s) from the matrix $\Xi_{\rm\bf zu}^{\rm\bf yv}(i,j)$ or the matrix $\Xi_{\rm\bf zv}^{\rm\bf zv}(i,j)$.

While the hypothesis adopted in Theorem \ref{theorem:5} and Corollary \ref{cor:2}, that is, $G_{\rm\bf zv}(\lambda,i)\equiv 0$ for each $i=1,2,\cdots,N$,
$G_{\rm\bf yv}\!(\lambda) \!=\! \overline{G}_{\rm\bf yv}\!(\lambda)G_{\rm\bf zv}(\lambda)$ and $G_{\rm\bf zu}\!(\lambda) \!=\! G_{\rm\bf zv}\!(\lambda)\overline{G}_{\rm\bf zu}\!(\lambda)$ with $\overline{G}_{\rm\bf yv}\!(\lambda)$ and $\overline{G}_{\rm\bf zu}\!(\lambda)$ are respectively of FNCR and FNRR, are quite restrictive, the corresponding NDS model still includes those adopted in \cite{cw2017,cw2020,ptt2019,vtc2021} as a special case, \textcolor{black}{noting that the models adopted in these works require that all subsystem outputs are measured, no matter it is an internal output or an external output. This requirement is equivalent to that in Equation (\ref{eqn:1}), $z(t,i)$ is identically equal to $y(t,i)$ for each $1\leq i \leq N$ and at each time constant $t$. On the other hand, the conditions of \cite{cw2017} and \cite{cw2020} are based on the TFM of the NDS, which may not be very helpful in subsystem selections and NDS designs, noting that this TFM is usually not available before system identification. In addition, computation complexity of these conditions increases rapidly with the number of NDS subsystems, which may make them less competitive in the analysis and synthesis of a large scale NDS.}

\textcolor{black}{ Assume now that
\begin{displaymath}
G_{\rm\bf yu}\!(\lambda) \!=\! S G_{\rm\bf zv}(\lambda)R, \hspace{0.15cm}
G_{\rm\bf yv}\!(\lambda) \!=\! S G_{\rm\bf zv}(\lambda), \hspace{0.15cm}
G_{\rm\bf zu}\!(\lambda) \!=\! G_{\rm\bf zv}\!(\lambda)R
\end{displaymath}
in which $S$ and $R$ are real matrices with compatible dimensions, while the TFMs $G_{\rm\bf yu}\!(\lambda)$, $G_{\rm\bf yv}\!(\lambda)$, $G_{\rm\bf zu}\!(\lambda)$ and $G_{\rm\bf zv}\!(\lambda)$ have the same definitions as those of Equation (\ref{eqn:6}). Then the NDS model adopted in this paper reduces to that of \cite{vtc2021}. Noting that the aforementioned assumptions can be satisfied through simply modifying the system matrices of each subsystem given in Equation (\ref{eqn:1}), it appears safe to declare that the NDS model of \cite{vtc2021} is a particular case satisfying Equation (\ref{eqn:34}) with both the TFM $\overline{G}_{\rm\bf yv}(\lambda)$ and the TFM $\overline{G}_{\rm\bf zu}(\lambda)$ being a constant real matrix. On the other hand, it can be easily understood that the condition of Corollary \ref{cor:2} is equivalent to the condition of \cite{vtc2021}.}

Nevertheless, further efforts are still required to remove the  hypotheses in Theorem \ref{theorem:5} and Corollary \ref{cor:2}, as they can not be easily satisfied in solving real world problems.

\section{Dependence on Subsystem Parameters}

In system designs, a first step is usually plant dynamics selection and parameter tuning. While Theorems \ref{theorem:2} and \ref{theorem:5}, as well as Corollary \ref{cor:2} of the previous section, clarify requirements on a subsystem such that the structure of an NDS constituted from it can be estimated from experiment data, it is still not clear how to select the dynamics and parameters of a plant to meet these requirements.

As pointed before, elements in system matrices of a plant are usually not algebraically independent of each other. Generally, these elements are functions of plant FPPs. In this section, we investigate relations between structure identifiability of an NDS and its subsystem (pseudo) FPPs, under the condition that system matrices of a subsystem are expressed as an LFT of its (pseudo) FPPs, as given in Equation (\ref{eqn:1-a}).

In order to study these relations, introduce auxiliary signal vectors $w(t,i)$ and $r(t,i)$ for each subsystem ${\rm\bf\Sigma}_{i}$ with $i=1,2,\cdots,N$, as follows,
\begin{eqnarray}
& & \hspace*{-1.5cm} w(t,i)= \left[\!\! \!\!\begin{array}{ccc}
F_{\rm\bf x}{(i)}\! & \!F_{\rm\bf v}{(i)}\! & \!F_{\rm\bf u}{(i)}
\end{array}\!\! \right]\left[\!\! \begin{array}{c}
{{x}(t,i)}\\
{{v}(t,i)}\\
{{u}(t,i)}
\end{array}\!\! \right] + G(i)r(t,i)
\label{eqn:7} \\
& & \hspace*{-1.5cm} r(t,i)=P(i)w(t,i)
\label{eqn:8}
\end{eqnarray}
With these auxiliary signal vectors, the dynamics of the $i$-th subsystem ${\rm\bf\Sigma}_{i}$, which are given by Equations (\ref{eqn:1}) and (\ref{eqn:1-a}), can be equivalently expressed as follows
\begin{equation}
\left[\!\! {\begin{array}{c}
{\delta({{x}}(t,i))}\\
{{w}(t,i)}\\
{{z}(t,i)}\\
{{y}(t,i)}
\end{array}}\!\! \right]  \!\!=\!\!
\left[\!\! \begin{array}{cccc}
{A^{[0]}_{\rm\bf xx}(i)} & H_{\rm\bf x}(i) & {A^{[0]}_{\rm\bf xv}(i)} & {B^{[0]}_{\rm\bf x}(i)}\\
F_{\rm\bf x}(i) & G(i) & F_{\rm\bf v}(i) & F_{\rm\bf u}(i) \\
{A^{[0]}_{\rm\bf zx}(i)} & H_{\rm\bf z}(i) & {A^{[0]}_{\rm\bf zv}(i)} & {B^{[0]}_{\rm\bf z}(i)}\\
{C^{[0]}_{\rm\bf x}(i)} & H_{\rm\bf y}(i) & {C^{[0]}_{\rm\bf v}(i)} & {D^{[0]}_{\rm\bf u}(i)}
\end{array}\!\! \right]\!\! \left[\!\! \begin{array}{c}
{{x}(t,i)}\\
r(t,i) \\
{{v}(t,i)}\\
{{u}(t,i)}
\end{array}\!\! \right]
\label{eqn:9}
\end{equation}

This approach has also been adopted in \cite{zz2020} to rewrite the subsystem model of an NDS with LFT parametrized subsystems in a form used in \cite{Zhou2015}, so that the results on NDS controllability/observability developed there can be applied. The purposes here, however, are completely different. Particularly, it is used to get an explicit expression respectively for the TFM  $G_{\rm\bf yv}(\lambda,i)$ and the TFM $G_{\rm\bf zu}(\lambda,i)$.

\textcolor{black}{On the other hand, it is worthwhile to mention that in its  subsystem dynamics and interaction descriptions, the above augmented system has completely the same form as the NDS of Equations (\ref{eqn:1}) and (\ref{eqn:2}).  Equations (\ref{eqn:8}) and (\ref{eqn:9}) also make it clear that if there are some (pseudo) FPPs of a subsystem that must be estimated from experiment data, then conclusions of the previous section can be directly utilized to verify its identifiability. The only required modifications are to include the parameters to be estimated into the SCM $\Phi$ of the augmented NDS.}

Taking Laplace/$\cal Z$ transformation on both sides of Equations (\ref{eqn:8}) and (\ref{eqn:9}) under the condition that the initial states of this subsystem are all equal to zero, the following equalities are obtained,
\begin{eqnarray}
& & \hspace*{-1.0cm} r(\lambda,i)=P(i)w(\lambda,i)
\label{eqn:10}\\
& & \hspace*{-1.0cm} \left[\!\!\! {\begin{array}{c}
{\lambda{{x}}(\lambda,i)}\\
{{w}(\lambda,i)}\\
{{z}(\lambda,i)}\\
{{y}(\lambda,i)}
\end{array}}\!\!\! \right]  \!\!=\!\!
\left[\!\!\! \begin{array}{cccc}
{A^{[0]}_{\rm\bf xx}(i)} & H_{\rm\bf x}(i) & {A_{\rm\bf xv}(i)} & {B^{[0]}_{\rm\bf x}(i)}\\
F_{\rm\bf x}(i) & G(i) & F_{\rm\bf v}(i) & F_{\rm\bf u}(i) \\
{A^{[0]}_{\rm\bf zx}(i)} & H_{\rm\bf z}(i) & {A^{[0]}_{\rm\bf zv}(i)} & {B^{[0]}_{\rm\bf z}(i)}\\
{C^{[0]}_{\rm\bf x}(i)} & H_{\rm\bf y}(i) & {C^{[0]}_{\rm\bf v}(i)} & {D^{[0]}_{\rm\bf u}(i)}
\end{array}\!\!\! \right]\!\!\! \left[\!\!\! \begin{array}{c}
{{x}(\lambda,i)}\\
r(\lambda,i) \\
{{v}(\lambda,i)}\\
{{u}(\lambda,i)}
\end{array}\!\!\! \right] \nonumber\\
\label{eqn:11}
\end{eqnarray}

Define TFMs $H_{\star\ddag}(\lambda,i)$ with $\star={\rm\bf w}$, ${\rm\bf z}$ or ${\rm\bf y}$, and $\ddag={\rm\bf r}$, ${\rm\bf v}$ or ${\rm\bf u}$, as follows,
\begin{eqnarray}
& &\hspace*{-0.4cm} \left[\!\! \begin{array}{ccc}
H_{\rm\bf wr}(\lambda,i) & H_{\rm\bf wv}(\lambda,i) & H_{\rm\bf wu}(\lambda,i) \\
H_{\rm\bf zr}(\lambda,i) & H_{\rm\bf zv}(\lambda,i) & H_{\rm\bf zu}(\lambda,i) \\
H_{\rm\bf yr}(\lambda,i) & H_{\rm\bf yv}(\lambda,i) & H_{\rm\bf yu}(\lambda,i) \\
\end{array}\!\! \right]  \nonumber\\
& & \hspace*{-0.6cm}=\!\!
\left[\!\!\! \begin{array}{ccc}
G(i) & F_{\rm\bf v}(i) & F_{\rm\bf u}(i) \\
H_{\rm\bf z}(i) & {A^{[0]}_{\rm\bf zv}(i)} & {B^{[0]}_{\rm\bf z}(i)}\\
H_{\rm\bf y}(i) & {C^{[0]}_{\rm\bf v}(i)} & {D^{[0]}_{\rm\bf u}(i)}  \end{array}\!\!\! \right]\!\!
+ \!\!\left[\!\! \begin{array}{c}
F_{\rm\bf x}(i)  \\
{A^{[0]}_{\rm\bf zx}(i)} \\
{C^{[0]}_{\rm\bf x}(i)} \end{array}\!\! \right]\!\!\left(\lambda I_{m_{{\rm\bf x}i}} - \right. \nonumber \\
& & \hspace*{1.2cm}
\left.{A^{[0]}_{\rm\bf xx}(i)}\right)^{-1}\!\!
\left[\!\! \begin{array}{ccc}
H_{\rm\bf x}(i) & {A^{[0]}_{\rm\bf xv}(i)} & {B^{[0]}_{\rm\bf x}(i)} \end{array}\!\! \right]
\label{eqn:12}
\end{eqnarray}
Then straightforward matrix operations prove that the relations among ${w}(\lambda,i)$, ${z}(\lambda,i)$, ${y}(\lambda,i)$, etc., which are given by Equation (\ref{eqn:11}), can be equivalently expressed as
\begin{equation}
\hspace*{-0.0cm} \left[\!\! {\begin{array}{c}
{{w}(\lambda,i)}\\
{{z}(\lambda,i)}\\
{{y}(\lambda,i)}
\end{array}}\!\! \right]  \!\!=\!\!
\left[\!\! \begin{array}{ccc}
H_{\rm\bf wr}(\lambda,i) & H_{\rm\bf wv}(\lambda,i) & H_{\rm\bf wu}(\lambda,i) \\
H_{\rm\bf zr}(\lambda,i) & H_{\rm\bf zv}(\lambda,i) & H_{\rm\bf zu}(\lambda,i) \\
H_{\rm\bf yr}(\lambda,i) & H_{\rm\bf yv}(\lambda,i) & H_{\rm\bf yu}(\lambda,i) \\
\end{array}\!\! \right] \!\!\!\!
\left[\!\! \begin{array}{c}
r(\lambda,i) \\
{{v}(\lambda,i)}\\
{{u}(\lambda,i)}
\end{array}\!\! \right]
\label{eqn:13}
\end{equation}

Let $m_{{\rm\bf w}i}$ stands for the dimension of the auxiliary signal vector $w(t,i)$. With similar arguments as those of Equations (\ref{eqn:15}) and (\ref{eqn:16}), it can be proven that if this subsystem is well-posed, which is equivalent to that the matrix $I-G(i)P(i)$ is invertible, then the TFM $I_{m_{{\rm\bf w}i}} \!-\! H_{\rm\bf wr}(\lambda,i)P(i)$ is not \textcolor{black}{identically} equal to zero. That is, its inverse is well defined. Combining Equations (\ref{eqn:10}) and (\ref{eqn:13}) together, direct algebraic manipulations show that the TFMs $G_{\rm\bf zv}(\lambda,i)$, $G_{\rm\bf zu}(\lambda,i)$, $G_{\rm\bf yv}(\lambda,i)$ and $G_{\rm\bf yu}(\lambda,i)$ of the previous section, can also be expressed as
\begin{eqnarray}
& &\hspace*{-0.8cm} \left[\!\! \begin{array}{cc}
G_{\rm\bf zv}(\lambda,i) & G_{\rm\bf zu}(\lambda,i)  \\
G_{\rm\bf yv}(\lambda,i) & G_{\rm\bf yu}(\lambda,i) \end{array}\!\! \right]
\!\! = \!\!
\left[\!\! \begin{array}{cc}
H_{\rm\bf zv}(\lambda,i) & H_{\rm\bf zu}(\lambda,i) \\
H_{\rm\bf yv}(\lambda,i) & H_{\rm\bf yu}(\lambda,i) \\
\end{array}\!\! \right] \!\! +   \nonumber\\
& & \hspace*{1.4cm}
\left[\!\! \begin{array}{c}
H_{\rm\bf zr}(\lambda,i)  \\
H_{\rm\bf yr}(\lambda,i) \end{array}\!\! \right]\!P(i)
\left[I_{m_{{\rm\bf w}i}} \!-\! H_{\rm\bf wr}(\lambda,i)P(i)\right]^{-1} \!\!\!\times\nonumber\\
& & \hspace*{3cm}\left[\!\! \begin{array}{cc}
H_{\rm\bf wv}(\lambda,i) & H_{\rm\bf wu}(\lambda,i) \end{array}\!\! \right]
\label{eqn:14}
\end{eqnarray}
That is, in the $i$-th subsystem ${\rm\bf\Sigma}_{i}$, all the TFMs including those from its internal inputs to its external/internal outputs, can be expressed as an LFT of the matrix constituted from its (pseudo) FPPs.

From this LFT expression, the following results are established for the TFM $G_{\rm\bf yv}(\lambda,i)$ being of FNCR.

\begin{theorem}
Assume that the $i$-th subsystem ${\rm\bf\Sigma}_{i}$ is well-posed. Then its TFM $G_{\rm\bf yv}(\lambda,i)$ is of FNCR, if and only if there exists a $\lambda\in {\cal C}$ such that the matrix pencil $M(\lambda,i)$ defined as follows is of FCR,
\begin{equation}
M(\lambda,i) \!\!=\!\! \left[\!\!\begin{array}{ccc}
\lambda I_{m_{{\rm\bf x}i}} \!-\! {A^{[0]}_{\rm\bf xx}(i)} & -{A^{[0]}_{\rm\bf xv}(i)} & -H_{\rm\bf x}(i) \\
{C^{[0]}_{\rm\bf x}(i)} & {C^{[0]}_{\rm\bf v}(i)}  & H_{\rm\bf y}(i)   \\
P(i)F_{\rm\bf x}(i)  &   P(i)F_{\rm\bf v}(i)   &  P(i)G(i) \!-\! I_{m_{{\rm\bf p}i}}  \end{array}\!\!\!\right]
\label{eqn:17}
\end{equation}
in which $m_{{\rm\bf p}i}$ stands for the number of the rows in the matrix $P(i)$ that are constructed from the (pseudo) FPPs of this subsystem.
\label{theorem:3}
\end{theorem}

The proof of the above theorem is provided in the appendix.

The matrix pencil $M(\lambda,i)$ in the above theorem has a form very similar to the matrix pencil $M(\lambda)$ of \cite{Zhou2015,zz2020} which is used for controllability/observability verification of an NDS. The conditions, however, are completely different. More precisely, in NDS controllability/observability verifications, the matrix pencil $M(\lambda)$ is required to be FCR at each $\lambda\in {\cal C}$. But the above theorem only asks for the existence of one particular $\lambda\in {\cal C}$, at which the matrix pencil $M(\lambda,i)$ is of FCR. On the other hand, some of the techniques developed in \cite{Zhou2015,zz2020} can be borrowed here to deal with NDS structure identifiability.

When the matrix $\left[\!\!\begin{array}{ccc}
{C^{[0]}_{\rm\bf x}(i)} & {C^{[0]}_{\rm\bf v}(i)}  & H_{\rm\bf y}(i)  \end{array}\!\!\!\right]$ is of FCR, it is obvious that at each $\lambda\in {\cal C}$, the matrix pencil $M(\lambda,i)$ is of FCR. That is, the TFM $G_{\rm\bf yv}(\lambda,i)$ is certainly of FNCR. Therefore, in the remaining of this section, we only investigate the situation in which this matrix is column rank deficient.
In this case, its right null space has nonzero elements and $\left[\!\!\begin{array}{ccc}
{C^{[0]}_{\rm\bf x}(i)} & {C^{[0]}_{\rm\bf v}(i)}  & H_{\rm\bf y}(i)  \end{array}\!\!\!\right]^{\perp}$ is not a zero vector.

Partition the matrix $\left[\!\!\begin{array}{ccc}
{C^{[0]}_{\rm\bf x}(i)} & {C^{[0]}_{\rm\bf v}(i)}  & H_{\rm\bf y}(i)  \end{array}\!\!\!\right]^{\perp}$ as
\begin{equation}
\left[\!\!\begin{array}{ccc}
{C^{[0]}_{\rm\bf x}(i)} & {C^{[0]}_{\rm\bf v}(i)}  & H_{\rm\bf y}(i)  \end{array}\!\!\!\right]^{\perp}={\rm\bf col}\!\left\{N_{\rm\bf x}(i),\; N_{\rm\bf v}(i),\;N_{\rm\bf w}(i)\right\}
\label{eqn:18}
\end{equation}
in which the sub-matrices $N_{\rm\bf x}(i)$, $N_{\rm\bf v}(i)$ and $N_{\rm\bf w}(i)$ respectively have $m_{{\rm\bf x}i}$, $m_{{\rm\bf v}i}$ and $m_{{\rm\bf p}i}$ rows. Then according to Lemma \ref{lemma:1}, the matrix pencil $M(\lambda,i)$ is of FCR at a particular $\lambda\in {\cal C}$, if and only if at this $\lambda$, the following matrix pencil $\overline{M}(\lambda,i)$ is of FCR,
\begin{eqnarray}
& & \hspace*{-0.8cm} \overline{M}(\lambda,i) \nonumber\\
& & \hspace*{-1.0cm}=\!\! \left[\!\!\!\begin{array}{ccc}
\lambda I_{m_{{\rm\bf x}i}} \!-\! {A^{[0]}_{\rm\bf xx}(i)} & -{A^{[0]}_{\rm\bf xv}(i)} & -H_{\rm\bf x}(i) \\
P(i)F_{\rm\bf x}(i)  &   P(i)F_{\rm\bf v}(i)   &  P(i)G(i) \!-\! I_{m_{{\rm\bf p}i}}  \end{array}\!\!\!\right]\!\!\!\left[\!\!\!\begin{array}{c} N_{\rm\bf x}(i) \\ N_{\rm\bf v}(i) \\ N_{\rm\bf w}(i) \end{array}\!\!\!\right] \nonumber\\
& & \hspace*{-1.0cm}=\!\!
\left[\!\!\begin{array}{c}
\lambda N_{\rm\bf x}(i) \!-\! \left[{A^{[0]}_{\rm\bf xx}(i)}N_{\rm\bf x}(i)
\!+\! {A^{[0]}_{\rm\bf xv}(i)}N_{\rm\bf v}(i) \!+\! H_{\rm\bf x}(i) N_{\rm\bf w}(i) \right]\\
P(i)\left[F_{\rm\bf x}(i)N_{\rm\bf x}(i) \!+\! F_{\rm\bf v}(i)N_{\rm\bf v}(i) \!+\!
G(i)N_{\rm\bf w}(i) \right] \!-\! N_{\rm\bf w}(i)  \end{array}\!\!\!\right] \nonumber\\
& &
\label{eqn:19}
\end{eqnarray}

To verify whether or not the matrix pencil $\overline{M}(\lambda,i)$ is of FNCR, the KCF of the matrix pencil $\lambda N_{\rm\bf x}(i) \!-\! \left[{A^{[0]}_{\rm\bf xx}(i)}N_{\rm\bf x}(i)
\!+\! {A^{[0]}_{\rm\bf xv}(i)}N_{\rm\bf v}(i) \!+\! H_{\rm\bf x}(i) N_{\rm\bf w}(i)\right]$ is utilized.

According to Lemma \ref{lemma:3}, there exist two invertible real matrices $U(i)$ and $V(i)$, some unique nonnegative integers $\xi^{[i]}_{\rm\bf H}$, $\zeta^{[i]}_{\rm\bf K}$, $\zeta^{[i]}_{{\rm\bf L}}$, $\zeta^{[i]}_{{\rm\bf N}}$, $\zeta^{[i]}_{{\rm\bf J}}$, $\xi^{[i]}_{\rm\bf L}(j)|_{j=1}^{\zeta^{[i]}_{{\rm\bf L}}}$ and $\xi^{[i]}_{\rm\bf J}(j)|_{j=1}^{\zeta^{[i]}_{{\rm\bf J}}}$, as well as some unique positive integers $\xi^{[i]}_{\rm\bf K}(j)|_{j=1}^{\zeta^{[i]}_{{\rm\bf K}}}$ and $\xi^{[i]}_{\rm\bf N}(j)|_{j=1}^{\zeta^{[i]}_{{\rm\bf N}}}$, such that
\begin{eqnarray}
& &\hspace*{-0.2cm} \lambda N_{\rm\bf x}(i) \!-\! \left[{A^{[0]}_{\rm\bf xx}(i)}N_{\rm\bf x}(i)
\!+\! {A^{[0]}_{\rm\bf xv}(i)}N_{\rm\bf v}(i) \!+\! H_{\rm\bf x}(i) N_{\rm\bf w}(i)\right]  \nonumber \\
&=& \hspace*{-0.2cm}U(i) K(\lambda,i) V(i)
\label{eqn:20}
\end{eqnarray}
in which
\begin{eqnarray}
K(\lambda,i)\!\!\!\!&=&\!\!\!\!
{\rm\bf diag}\!\left\{\!L_{\xi^{[i]}_{\rm\bf L}(j)}(\lambda)|_{j=1}^{\zeta^{[i]}_{{\rm\bf L}}},\;H_{\xi^{[i]}_{{\rm\bf H}}}(\lambda),\;K_{\xi^{[i]}_{\rm\bf K}(j)}(\lambda)|_{j=1}^{\zeta^{[i]}_{{\rm\bf K}}}, \right.\nonumber\\
& & \hspace*{1.5cm}\left. N_{\xi^{[i]}_{\rm\bf N}(j)}(\lambda)|_{j=1}^{\zeta^{[i]}_{{\rm\bf N}}},\; J_{\xi^{[i]}_{\rm\bf J}(j)}(\lambda)\!|_{j=1}^{\zeta^{[i]}_{{\rm\bf J}}}\!\right\}
\label{eqn:21}
\end{eqnarray}

From this KCF and Lemma \ref{lemma:2}, the following results are obtained, while their proof is deferred to the appendix.

\begin{corollary}
Define matrices $\Theta(i)$ and $\Pi(i)$ respectively as
\begin{eqnarray*}
& & \hspace*{-1.0cm}\Theta(i)\!=\!\left[F_{\rm\bf x}(i)N_{\rm\bf x}(i) \!+\! F_{\rm\bf v}(i)N_{\rm\bf v}(i) \!+\! G(i)N_{\rm\bf w}(i) \right]\!V^{-1}(i,{\rm\bf m}(i)) \\
& & \hspace*{-1.0cm}\Pi(i)\!=\! N_{\rm\bf w}(i)V^{-1}(i,{\rm\bf m}(i))
\end{eqnarray*}
in which ${\rm\bf m}(i)={\zeta^{[i]}_{{\rm\bf L}}} + \sum_{j=1}^{\zeta^{[i]}_{{\rm\bf L}}}{\xi^{[i]}_{\rm\bf L}(j)}$, while $V^{-1}(i,{\rm\bf m}(i))$ is the sub-matrix of the inverse of the matrix $V(i)$ consisting of its first ${\rm\bf m}(i)$ columns. Then the matrix pencil   $\overline{M}(\lambda,i)$ is of FNCR, if and only if the following matrix pencil $\tilde{M}(\lambda,i)$ is of FNCR,
\begin{equation}
\tilde{M}(\lambda,i)=\left[\!\!\begin{array}{c}
{\rm\bf diag}\!\left\{\!L_{\xi^{[i]}_{\rm\bf L}(j)}(\lambda)|_{j=1}^{\zeta^{[i]}_{{\rm\bf L}}} \right\}
 \\
P(i)\Theta(i)-\Pi(i)  \end{array}\!\!\right]
\label{eqn:23}
\end{equation}
\label{cor:1}
\end{corollary}

Compared with the matrix pencil $\overline{M}(\lambda,i)$, the matrix pencil $\tilde{M}(\lambda,i)$ usually has much less columns. This means that the condition of Corollary \ref{cor:1} is in general much more computationally attractive than that of Theorem \ref{theorem:3}. On the other hand, from the proof of the above corollary, it is also clear that if $\zeta^{[i]}_{{\rm\bf L}}=0$, that is, if there does not exist a matrix pencil in the form of $L_{\star}(\lambda)$ in the KCF of the matrix pencil $\lambda N_{\rm\bf x}(i) \!-\! \left[{A^{[0]}_{\rm\bf xx}(i)}N_{\rm\bf x}(i)
\!+\! {A^{[0]}_{\rm\bf xv}(i)}N_{\rm\bf v}(i) \!+\! H_{\rm\bf x}(i) N_{\rm\bf w}(i)\right]$, then the matrix pencil $\overline{M}(\lambda,i)$, and therefore the TFM
$G_{\rm\bf yv}(\lambda,i)$, is certainly of FNCR.

To establish a more direct and computationally attractive condition on subsystem dynamics and (pseudo) FPPs, partition the matrix $\Theta(i)$ and the matrix $\Pi(i)$ respectively as
\begin{eqnarray}
& & \Theta(i)=\left[\Theta_{1}(i) \;\; \Theta_{2}(i)\;\; \cdots \;\; \Theta_{\zeta^{[i]}_{{\rm\bf L}}}(i)   \right]  \\
& & \Pi(i)=\left[\Pi_{1}(i) \;\; \Pi_{2}(i)\;\; \cdots \;\; \Pi_{\zeta^{[i]}_{{\rm\bf L}}}(i)   \right]
\end{eqnarray}
Here, for each $j=1,2,\cdots, \zeta^{[i]}_{{\rm\bf L}}$, both the sub-matrix $\Theta_{j}(i)$ and the sub-matrix $\Pi_{j}(i)$ have
$\xi^{[i]}_{\rm\bf L}(j)+1$ columns. Define a positive integer $\xi^{[i]}_{\rm\bf L}$ as
\begin{equation}
\xi^{[i]}_{\rm\bf L}=\max_{j\in\left\{1,2,\cdots,  \zeta^{[i]}_{{\rm\bf L}}\right\}} \xi^{[i]}_{\rm\bf L}(j)
\end{equation}
Moreover, for every $j$ belonging to the set $\left\{1,2,\cdots, \zeta^{[i]}_{{\rm\bf L}}\right\}$, define a matrix $\overline{\Theta}_{j}(i)$ and a matrix $\overline{\Pi}_{j}(i)$ respectively through
\begin{equation}
\overline{\Theta}_{j}(i)=\left[\Theta_{j}(i)\;\; 0\right], \hspace{0.5cm}
\overline{\Pi}_{j}(i)=\left[\Pi_{j}(i)\;\; 0\right]
\end{equation}
so that all of them have $\xi^{[i]}_{\rm\bf L}+1$ columns.

Using the above symbols, on the basis of the structure of the null space of a matrix pencil with the form $L_{\star}(\lambda)$, the following conditions are derived for the TFM $G_{\rm\bf yv}(\lambda,i)$ to be FNCR. These conditions are computationally more attractive, give more direct requirements on subsystem dynamics and (pseudo) FPPs, and therefore may be more insightful in selecting subsystem dynamics and parameters.

\begin{theorem}
Define MVPs $\Theta(\lambda,i)$ and $\Pi(\lambda,i)$ respectively as
\begin{eqnarray*}
& &
\hspace*{-0.5cm}\Theta(\lambda,i)\!=\!\left[
\Theta_{1}(i){\rm\bf col}\!\left\{\!\!\left.\lambda^{k}\right|_{k=0}^{\xi^{[i]}_{\rm\bf L}(1)}\!\!\right\}\;\; \cdots \;\;
\Theta_{\zeta^{[i]}_{{\rm\bf L}}}(i){\rm\bf col}\!\left\{\!\!\left.\lambda^{k}\right|_{k=0}^{\xi^{[i]}_{\rm\bf L}(\zeta^{[i]}_{{\rm\bf L}})}\!\!\right\} \!\! \right]  \\
& &
\hspace*{-0.5cm}\Pi(\lambda,i)\!=\!\left[
\Pi_{1}(i){\rm\bf col}\!\left\{\!\!\left.\lambda^{k}\right|_{k=0}^{\xi^{[i]}_{\rm\bf L}(1)}\!\!\right\}\;\; \cdots \;\;
\Pi_{\zeta^{[i]}_{{\rm\bf L}}}(i){\rm\bf col}\!\left\{\!\!\left.\lambda^{k}\right|_{k=0}^{\xi^{[i]}_{\rm\bf L}(\zeta^{[i]}_{{\rm\bf L}})}\!\!\right\} \!\! \right]
\end{eqnarray*}
Moreover, define a matrix $\Gamma(i)$ as
\begin{eqnarray*}
\Gamma(i) \!\!\!\!&=&\!\!\!\! \left[(\overline{\Theta}_{1}(i)\otimes I){\rm\bf vec}(P(i))-{\rm\bf vec}(\Pi_{1}(i))\;\; \cdots \right.\\
& & \hspace*{1cm}
\left.\!\! (\overline{\Theta}_{\zeta^{[i]}_{{\rm\bf L}}}(i)\otimes I){\rm\bf vec}(P(i))\!-\!{\rm\bf vec}\!\left(\Pi_{\zeta^{[i]}_{{\rm\bf L}}}(i)\right)\!\right]
\end{eqnarray*}
Then
\begin{itemize}
\item the TFM $G_{\rm\bf yv}(\lambda,i)$ is of FNCR, if and only if the MVP  $P(i)\Theta(\lambda,i)-\Pi(\lambda,i)$ is.
\item the TFM $G_{\rm\bf yv}(\lambda,i)$ is of FNCR, only if the matrix  $\Gamma(i)$ is of FCR.
\end{itemize}
\label{theorem:4}
\end{theorem}

The proof of this theorem is also deferred to the appendix.

Using similar arguments as those between Equation (\ref{eqn:a27}) to (\ref{eqn:a31}) in the proof of Corollary \ref{cor:1}, it can be proven that the MVP $P(i)\Theta(\lambda,i)-\Pi(\lambda,i)$ is of FNCR, if and only if its Smith form has the following structure
\begin{displaymath}
U(\lambda,i)
\!\left[\!\begin{array}{c}
{\rm\bf diag}\left\{\left.{\alpha}^{[j]}(\lambda,i)\right|_{j=1}^{\zeta^{[i]}_{{\rm\bf L}}}\right\}  \\
0  \end{array}\!\right]\!
V(\lambda,i)
\end{displaymath}
in which both $U(\lambda,i)$ and $V(\lambda,i)$ are unimodular matrices with a compatible dimension, while  ${\alpha}^{[j]}(\lambda,i)|_{j=1}^{\zeta^{[i]}_{{\rm\bf L}}}$ are some
nonzero and real coefficient polynomials with a finite degree. The latter can be verified through various standard methods developed in matrix analysis, system analysis and synthesis, etc. \cite{Kailath1980, zdg1996}.

Note that
\begin{equation}
P(i)\Theta(\lambda,i)-\Pi(\lambda,i)=\left[\begin{array}{cc} P(i) & -I \end{array}\right]
\left[\begin{array}{c} \Theta(\lambda,i) \\ \Pi(\lambda,i) \end{array}\right]
\label{eqn:28}
\end{equation}
It is obvious that the MVP $P(i)\Theta(\lambda,i)-\Pi(\lambda,i)$ is of FNCR, only if the MVP ${\rm\bf col}\!\left\{\Theta(\lambda,i),\; \Pi(\lambda,i)\right\}$ is. As the latter is independent of the subsystem parameters, it gives conditions on subsystem dynamics such that \textcolor{black}{an NDS can be constituted from it with the NDS structure being identifiable.}

The 2nd condition of Theorem \ref{theorem:4} depends affinely on subsystem (pseudo) FPPs, which is helpful in understanding influences of these parameters on NDS structure identifiability.

Note that the TFM $G_{\rm\bf zu}(\lambda,i)$ is of FNRR, if and only if the TFM $G^{T}_{\rm\bf zu}(\lambda,i)$ is of FNCR. This means that the above results for verifying whether or not the TFM $G_{\rm\bf yv}(\lambda,i)$ is of FNCR, can also be applied to verify whether or not the TFM $G_{\rm\bf zu}(\lambda,i)$ is of FNRR. The details are omitted due to their obviousness and close similarities.

While it is interesting to derive conditions on subsystem dynamics and parameters from Theorem \ref{theorem:5} and Corollary \ref{cor:2} for NDS structure identifiability, some mathematical difficulties are still to be settled.

\section{A Numerical Example}

To illustrate theoretical results obtained in the previous sections, an artificial NDS is constructed in this section which has 2 subsystems and each of whose subsystems consists of 3 capacitors, 6 operation amplifiers, 2 varistors and 18 resistors, as shown in Figure \ref{fig:1}. To make presentation concise, it is assumed that in each subsystem, all the capacitors take the same value $C_{i}$, while all the resistors take the same value $R_{i}$. In addition, the maximum values of the varistors, that is, $R_{i1}$ and $R_{i2}$, are assumed to be $R_{i}$.
Here, $i\in\{1,2\}$ stands for the subsystem index.

\renewcommand{\thefigure}{\arabic{figure}}
\setcounter{figure}{0}
\vspace{-0.0cm}
\begin{figure}[!ht]
\vspace{-0.0cm}
\begin{center}
\includegraphics[width=3.0in]{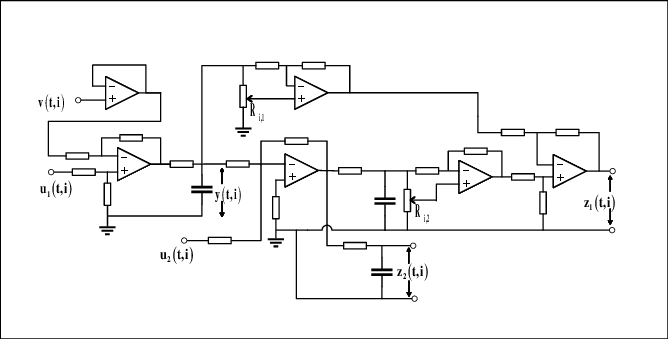}
\vspace{-0.5cm}\hspace*{5cm} \caption{The $i$-th Subsystem of the Artificial NDS}
\label{fig:1}
\end{center}
\end{figure}
\vspace{-0.0cm}

Define pseudo FPPs $T_{i}$, $k_{i1}$ and $k_{i2}$ of the $i$-th subsystem ${\rm\bf\Sigma}_{i}$ with $i=1,2$, respectively as
\begin{displaymath}
T_{i}=R_{i}C_{i},\hspace{0.25cm} k_{i1}=\frac{R_{i1}}{R_{i}},\hspace{0.25cm} k_{i2}=\frac{R_{i2}}{R_{i}}
\end{displaymath}
To investigate influences of the resistors $R_{i1}$ and $R_{i2}$ on NDS structure identifiability, assume that the time constant $T_{i}$ is prescribed for each subsystem.

Take the voltage of each capacitor as a state of a subsystem, while the voltage of the leftmost capacitor as its output. Under the above assumptions, the system matrices of each subsystem can be written as follows,
\begin{eqnarray*}
& & \hspace*{-0.7cm}
A_{\rm\bf xx}^{[0]}(i) \!=\! -\frac{1}{T_{i}}\!\!\left[\!\!\! \begin{array}{ccc}
3 & 0 & 0 \\ 1 & 2 & 0 \\ 1 & 0 & 1
\end{array}\!\!\! \right]\!\!, \hspace{0.3cm}
A_{\rm\bf xv}^{[0]}(i) \!=\! -\frac{1}{T_{i}}\!\!\left[\!\!\! \begin{array}{c}
1  \\ 0 \\ 0
\end{array}\!\!\! \right]  \\
& & \hspace*{-0.7cm}
B_{\rm\bf x}^{[0]}(i) \!=\! -\frac{1}{T_{i}}\!\!\left[\!\!\! \begin{array}{cc}
-1 & 0  \\ 0 & 1 \\  0 & 1
\end{array}\!\!\! \right]\!\!, \hspace{0.3cm}
A_{\rm\bf zx}^{[0]}(i) \!=\! \!\left[\!\!\! \begin{array}{ccc}
-1 & 1 & 0  \\ 0 & 0 & 1
\end{array}\!\!\! \right]  \\
& & \hspace*{-0.7cm}
A_{\rm\bf zv}^{[0]}(i)=0_{2},\hspace{0.3cm}
B_{\rm\bf z}^{[0]}(i)=0_{2\times 2}  \\
& & \hspace{-0.7cm}
C_{\rm\bf x}^{[0]}(i)=[1\;\; 0\;\; 0],\hspace{0.2cm}
C_{\rm\bf v}^{[0]}(i)=0,\hspace{0.2cm}
D_{\rm\bf u}^{[0]}(i)=[0\;\; 0]   \\
& & \hspace*{-0.7cm}
H_{\rm\bf x}(i)\!=\! -\frac{1}{T_{i}}\!\!\left[\!\!\! \begin{array}{cc}
1 & 0 \\ 0 & 1 \\ 0 & 0
\end{array}\!\!\! \right]\!\!, \hspace{0.2cm}
H_{\rm\bf z}(i)\!=\! \!\left[\!\!\! \begin{array}{cc}
2 & -2 \\ 0 & 0
\end{array}\!\!\! \right]\!\!, \hspace{0.2cm}
H_{\rm\bf y}(i)\!=\! [0\;\;0] \\
& & \hspace*{-0.7cm}
F_{\rm\bf x}(i)\!=\!\!\left[\!\!\! \begin{array}{ccc}
1 & 0 & 0\\ 0 & 1 & 0
\end{array}\!\!\! \right]\!\!, \hspace{0.1cm}
F_{\rm\bf v}(i)\!=\! 0_{2}, \hspace{0.1cm}
F_{\rm\bf u}(i)\!=\! 0_{2\times 2}, \hspace{0.1cm} G(i)\!=\! I_{2}
\end{eqnarray*}
Moreover,
\begin{displaymath}
P(i) \!=\! \!\!\left[\!\!\! \begin{array}{cc}
\frac{k_{i1}}{k_{i1}+1} & 0 \\ 0 & \frac{k_{i2}}{k_{i2}+1}
\end{array}\!\!\! \right]
\end{displaymath}

From these system matrices, direct algebraic manipulations show that for each $i=1,2$, $G_{\rm\bf yv}(\lambda,i)$ is a nonzero $1\times 1$ dimensional transfer function, while the TFM $G_{\rm\bf zu}(\lambda,i)$ has the following expression,
\begin{displaymath}
G_{\rm\bf zu}(\lambda,\!i) \!\!= \!\!\left[\!\!\!\! \begin{array}{cc}
\frac{T_{i}(2k_{i1}-1)}{\lambda T_{i}+k_{i1}+3} \!+\!
\frac{2k_{i2}-1}{(\lambda T_{i}+k_{i1}+3)(\lambda T_{i}+k_{i2}+2)}
 & \frac{2k_{i2}-1}{\lambda T_{i}+k_{i2}+2} \\
\frac{-1}{(\lambda T_{i}+1)(\lambda T_{i}+k_{i1}+3)}
& \frac{-1}{\lambda T_{i}+1}
\end{array}\!\!\!\! \right]
\end{displaymath}
We therefore have that
\begin{displaymath}
{\rm\bf det}\left\{G_{\rm\bf zu}(\lambda,i)\right\} = - \frac{T_{i}(2k_{i1}-1)}{(\lambda T_{i}+1)(\lambda T_{i}+k_{i1}+3)}
\end{displaymath}
Note that the time constant $T_{i}$ always takes a positive value. It is clear that ${\rm\bf det}\left\{G_{\rm\bf zu}(\lambda,i)\right\} \equiv 0$, if and only if
$k_{i1}=0.5$.

These observations mean that for each $i=1,2$, the TFM $G_{\rm\bf yv}(\lambda,i)$ is always of FNCR, while the TFM $G_{\rm\bf zu}(\lambda,i)$ is not of FNRR when and only when $k_{i1}$ takes the value $0.5$.

As relations have already become clear between the pseudo FPPs $k_{i1}/k_{i2}$ and the required properties that the TFM $G_{\rm\bf yv}(\lambda,i)$ is of FNCR/the TFM $G_{\rm\bf zu}(\lambda,i)$ is of FNRR, results of Theorem \ref{theorem:4}, etc., are no longer required for this example.

It can therefore be declared from Theorem \ref{theorem:2} that if each of the NDS subsystems ${\rm\bf\Sigma}_{1}$ and ${\rm\bf\Sigma}_{2}$ has its pseudo FPP $k_{11}/k_{21}$ taking a value different from $0.5$, then \textcolor{black}{the structure of this artificial NDS is identifiable.}

To investigate the influences of the pseudo FPPs $k_{11}$ and $k_{21}$ on the structure identifiability of the artificial NDS, fix the other pseudo FPPs as $T_{1}=T_{2}=1s$, $k_{12}=0.4$ and $k_{22}=0.9$. Moreover, the pseudo FPPs $k_{11}$ and $k_{21}$ are assumed to take the same value $k_{1}$ that varies between $0$ and $1$. For each value of $k_{1}$, the following quantity $d_{sid}^{\rm [F]}$ is adopted as a  frequency domain measure for the distance of an NDS to the set of NDSs \textcolor{black}{with their structures unidentifiable,}
\begin{displaymath}
 d_{sid}^{\rm [F]} =  \inf_{\Phi_{1}}\inf_{\Phi_{2}\neq \Phi_{1}} \frac{\left|\left|H(\lambda,\Phi_{2})-H(\lambda,\Phi_{1})\right|\right|_{\infty}}{
 \overline{\sigma}(\Phi_{2}-\Phi_{1})}
\end{displaymath}
in which $||\cdot||_{\infty}$ stands for the ${\cal L}_{\infty}$-norm of a TFM, while $\overline{\sigma}(\cdot)$ the maximum singular value of a matrix. Moreover, the TFM $H(\lambda,\Phi_{1})$ and the TFM $H(\lambda,\Phi_{2})$ have a definition given  by Equation (\ref{eqn:6})

Note that the ${\cal L}_{\infty}$-norm of a TFM is defined as the supremum of the maximum singular value of its frequency response, which is well known to be an induced norm \cite{zdg1996}. On the other hand, the maximum singular value of a matrix is also an induced norm that is extensively adopted in matrix difference measurements \cite{Gantmacher1959,hj1991}. The above quantity appears reasonable from an application viewpoint. Another option is to use the Frobenius norm of a matrix in the definition of the above quantity $d_{sid}^{\rm [F]}$. Our simulation results, however, show that the conclusions are almost the same.

While from an application viewpoint, the above quantity may be a good candidate in measuring the distance of an NDS \textcolor{black}{to the set of NDSs whose structures are unidentifiable}, there are currently still some mathematical difficulties to be settled for calculating it analytically. It is therefore estimated in this section through numerical simulations.

Specifically, for a prescribed value of the parameter $k_{1}$, $10^{3}$ SCM $\Phi_{1}$s are randomly and independently generated. In addition, for each randomly generated SCM $\Phi_{1}$, $2.0000\times 10^{3}$ SCM $\Phi_{2}$s are generated, also randomly and independently.
Furthermore, for each generated SCM pair $\Phi_{1}$ and $\Phi_{2}$, the ${\cal L}_{\infty}$-norm of the TFM $H(\lambda,\Phi_{2})-H(\lambda,\Phi_{1})$ is divided by the maximum singular value of the matrix $\Phi_{2}-\Phi_{1}$. Finally, the minimum value of this division over all the generated SCM $(\Phi_{1},\;\Phi_{2})$ pairs is used as an estimate for the above frequency domain  measure $d_{sid}^{\rm [F]}$ on the distance of an NDS \textcolor{black}{to the set of NDSs whose structure is unidentifiable}, while the SCM pair $\Phi_{1}$ and $\Phi_{2}$ associated with this minimum value are recorded respectively as $\Phi_{1}^{[k_{1}]}$ and $\Phi_{2}^{[k_{1}]}$.

In the aforementioned SCM generation, each element of the SCM is produced independently, according to the continuous uniform distribution over the interval $(-1,\; 1)$. 

\vspace{-0.0cm}
\begin{table}[t]
\vspace*{-0.00cm}
\caption{\vspace*{-0.00cm} Estimated NDS Distance \textcolor{black}{to the Set of NDSs with an Unidentifiable Structure}}
\begin{center}
\vspace{-0.2cm}
\begin{tabular}{|c||c|c|c|}
\hline
$k_{1}$  & $d_{scm}$ & $d_{sid}^{\rm [F]}$  & $d_{sid}^{\rm [T]}$ \\
\hline
\hline
$0.0500$ & 1.6138 &	$5.7399\times 10^{-2}$  & $7.0091\times 10^{-5}$ \\ \hline
$0.1000$ & 1.2643 &	$5.2766\times 10^{-2}$  & $7.1263\times 10^{-5}$ \\ \hline
$0.1500$ & $7.7771\times 10^{-1}$ &	$4.0053\times 10^{-2}$  & $5.5975\times 10^{-5}$ \\ \hline
$0.2000$ & 1.2256 &	$4.0338\times 10^{-2}$  & $5.2004\times 10^{-5}$ \\ \hline
$0.2500$ & 1.9969 &	$3.4435\times 10^{-2}$  & $3.9901\times 10^{-5}$ \\ \hline
$0.3000$ & 2.0254 &	$3.2441\times 10^{-2}$  & $4.0897\times 10^{-5}$ \\ \hline
$0.3500$ & 1.2933 &	$2.5469\times 10^{-2}$  & $3.3994\times 10^{-5}$ \\ \hline
$0.4000$ & 2.1485 &	$1.9829\times 10^{-2}$  & $2.8705\times 10^{-5}$ \\ \hline
$0.4500$ & 1.8778 &	$1.4281\times 10^{-2}$  & $1.8896\times 10^{-5}$ \\ \hline
$0.5000$ & 1.7358 &	$1.2580\times 10^{-2}$  & $1.5758\times 10^{-5}$ \\ \hline
$0.5500$ & 1.7297 &	$1.1353\times 10^{-2}$  & $1.6384\times 10^{-5}$ \\ \hline
$0.6000$ & 2.2971 &	$1.4394\times 10^{-2}$  & $2.0664\times 10^{-5}$ \\ \hline
$0.6500$ & 1.3653 &	$2.0464\times 10^{-2}$  & $3.2608\times 10^{-5}$ \\ \hline
$0.7000$ & 1.8584 &	$2.2381\times 10^{-2}$  & $3.0936\times 10^{-5}$ \\ \hline
$0.7500$ & 1.7534 &	$2.9100\times 10^{-2}$  & $3.6359\times 10^{-5}$ \\ \hline
$0.8000$ & 2.1193 &	$3.1190\times 10^{-2}$  & $4.4965\times 10^{-5}$ \\ \hline
$0.8500$ & 1.2331 &	$3.0762\times 10^{-2}$  & $5.0705\times 10^{-5}$ \\ \hline
$0.9000$ & 1.4441 &	$3.4366\times 10^{-2}$  & $5.1064\times 10^{-5}$ \\ \hline
$0.9500$ & 1.8608 &	$3.8835\times 10^{-2}$  & $4.9123\times 10^{-5}$ \\ \hline
\end{tabular}
\vspace{-0.5cm}
\end{center}
\label{table:1}
\end{table}
\vspace{-0.0cm}

On the other hand, it is invaluable from an application viewpoint to see the maximum difference in the time domain between the outputs of two artificial NDSs that are closest to each other in some sense and associated with different SCMs, with sampled input/output data of a finite length. However, exhausting all the possible probing signals are computationally prohibitive. Instead, a pseudo random binary signal (PRBS) is  adopted as testing signals, which has been extensively utilized in traditional system identifications and is widely recognized as an informative probing signal that is persistently exciting \cite{vdhb2013,hgb2019}. \textcolor{black}{Note that it is observed in \cite{wrmd2005} that all the input/output trajectories of a system can be expressed as a linear combination of the input/output trajectory of the system stimulated by a persistently exciting signal. This probing signal selection appears reasonable.}

\vspace{-0.0cm}
\begin{figure}[!ht]
\begin{center}
\includegraphics[width=3.0in]{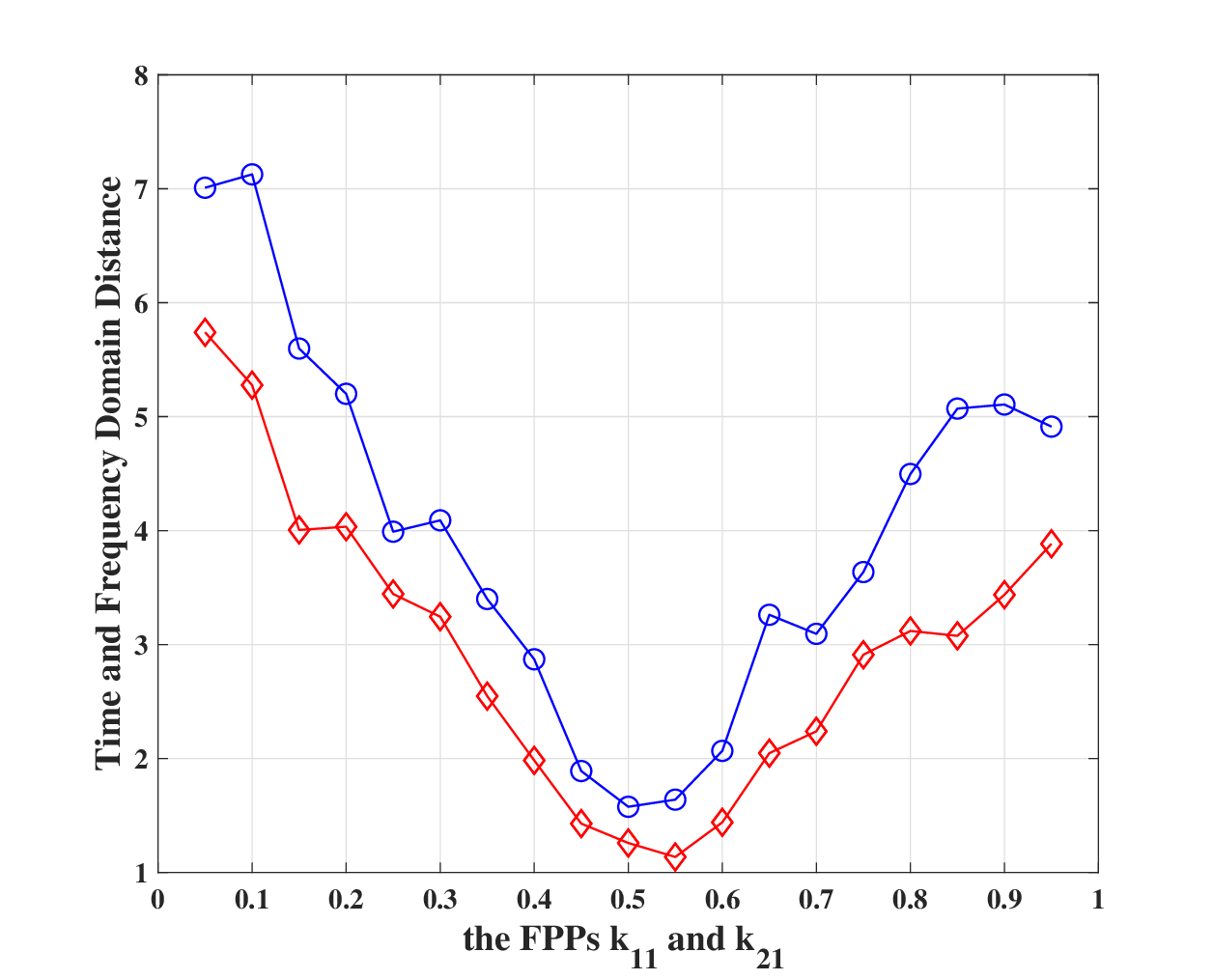}
\vspace{-0.5cm}\hspace*{5cm} \caption{Normalized Estimates of the Distance of the NDS \textcolor{black}{to the Set of NDSs with an Unidentifiable Structure}. $\Diamond$: frequency domain distance, $\circ$: time domain distance.}
\label{fig:2}
\end{center}
\end{figure}
\vspace{-0.0cm}

Particularly, for each value of $k_{1}$, stimulate simultaneously the NDS ${\rm\bf\Sigma}\left(\Phi_{1}^{[k_{1}]}\right)$ and the NDS ${\rm\bf\Sigma}\left(\Phi_{2}^{[k_{1}]}\right)$ with 4 independent PRBSs that take values from the set $\{-1,\;1\}$. Recall that $\Phi_{1}^{[k_{1}]}$ and $\Phi_{2}^{[k_{1}]}$  stand for the SCM pair among all the randomly generated samples, that make the cost function $\frac{\left|\left|H(\lambda,\Phi_{2})-H(\lambda,\Phi_{1})\right|\right|_{\infty}}{ \overline{\sigma}(\Phi_{2}-\Phi_{1})}$ achieve its minimum value. This means that the aforementioned two NDSs are the closest ones among all the sampled NDS pairs. The differences $e(t)|_{t=0}^{M-1}$ between the outputs of these two artificial NDSs are measured by their extended $l_{2}$-norm $||e(t)||_{2}$ defined as $||e(t)||_{2}=\sqrt{\sum_{t=0}^{M-1}e^{T}(t)e(t)}$, in which $e(t) = y\left(t,\Phi_{1}^{[k_{1}]}\right) - y\left(t,\Phi_{2}^{[k_{1}]}\right)$ stands for the output difference at the $t$-th sampling instant, while $M$ is the number of all sampled time instants. With this information available, the following quantity is calculated,
\begin{displaymath}
\frac{\left|\left|e(t)\right|\right|_{2}}{M \overline{\sigma}\left(\Phi_{2}^{[k_{1}]}-\Phi_{1}^{[k_{1}]}\right)}
\end{displaymath}
This quantity is referred here to as an estimate of a time domain measure for the distance of an NDS to the set of \textcolor{black}{NDSs whose structures are unidentifiable,} which is denoted by $d_{sid}^{\rm [T]}$.

For each prescribed value of $k_{1}$, let $A(k_{1})$ stand for the block diagonal matrix constituted from the state transition matrices of the two associate NDSs ${\rm\bf\Sigma}\left(\Phi_{1}^{[k_{1}]}\right)$ and ${\rm\bf\Sigma}\left(\Phi_{2}^{[k_{1}]}\right)$, while $\rho_{max}(A(k_{1}))$ and $\rho_{min}(A(k_{1}))$ respectively the maximum and minimum of the absolute value of its eigenvalues. In actual simulations, an NDS is sampled with a constant period set to one-tenth of $\frac{1}{\rho_{max}(A(k_{1}))}$, while the sample number to the maximum of $2.0000\times 10^{4}$ and $100\times\frac{\rho_{max}(A(k_{1}))}{\rho_{min}(A(k_{1}))}$. This simulation setting is adopted to get sufficient information about the required output differences without too much computation costs.

The computation results are listed in Table \ref{table:1} for both the frequency and the time domain distances. For reference, $\overline{\sigma}\left(\Phi_{2}^{[k_{1}]}-\Phi_{1}^{[k_{1}]}\right)$, that is, the maximum singular value of the difference between the associated two SCMs which lead to the $d_{sid}^{\rm [F]}$ estimate for the corresponding value of $k_{1}$, are also given there. It is denoted by $d_{scm}$ in the table.

Obviously, when $k_{11}$ and $k_{21}$ take a value around $0.5000$, both the $d_{sid}^{\rm [F]}$ and the $d_{sid}^{\rm [T]}$ reach its minimum value. In addition, while neither of them is equal to zero there, both of them are much closer to zero than their counterparts when $k_{11}$ and $k_{21}$ take other values. This means that although $k_{11}=k_{21}=0.5000$ does not lead to \textcolor{black}{an NDS whose structure is  unidentifiable}, this NDS is much closer to \textcolor{black}{the set of NDSs with an unidentifiable structure} than an NDS when these two pseudo FPPs are far away from $0.5000$. This may also imply that while Theorem \ref{theorem:2} only gives a sufficient condition for \textcolor{black}{the structure of an NDS to be identifiable}, satisfaction of this condition may make the structure of an NDS much easier to be estimated. However, further efforts are still required to give a rigorous theoretical guarantee.  On the other hand, for all the computed $k_{1}$ including $0.5000$, the value of $d_{scm}$ is not very close to zero, which is equivalent to that the SCM $\Phi_{1}^{[k_{1}]}$ and the SCM $\Phi_{2}^{[k_{1}]}$ are not very close to each other. It can therefore be declared that two SCMs far away from each other are still possible to generate similar NDS input-output relations. This may not be out of imaginations, noting that the TFM of the NDS $\rm\bf\Sigma$ can be expressed as an LFT of its SCM, while an LFT is in fact a nonlinear transformation.

To see the variation trends of $d_{sid}^{\rm [F]}$ and $d_{sid}^{\rm [T]}$ more clearly when the FPPs $k_{11}$ and $k_{21}$ take different values, the computed numbers are also plotted in Figure 2. Here, the estimated $d_{sid}^{\rm [F]}$ and $d_{sid}^{\rm [T]}$ are respectively normalized by a factor $10^{2}$ and a factor $10^{5}$ with the purposes of improving figure readability.

Clearly, both the estimated $d_{sid}^{\rm [F]}$ and the estimated $d_{sid}^{\rm [T]}$ increase almost monotonically with the magnitude increment of the deviation of the pseudo FPPs $k_{11}$ and $k_{21}$ from $0.5000$. On the other hand, their curves have very similar variation patterns. This may suggest that these two indices are consistent well with each other, and may be good candidates for a metric in measuring distance of an NDS \textcolor{black}{to the set of NDSs with an unidentifiable structure.}

\section{Concluding Remarks}

In this paper, we have investigated conditions on a subsystem such that a linear time invariant NDS constructed from it \textcolor{black}{has an identifiable structure}, that is, the subsystem interactions can be estimated from experiment data. Except well-posedness, there are neither any other restrictions on subsystem dynamics, nor any other restrictions on subsystem connections. It is proven that \textcolor{black}{the structure of an LTI NDS is identifiable}, if the TFMs of its subsystems meet some rank conditions. Based on this result, it has been further shown that in order to guarantee the satisfaction of this condition, it is necessary and sufficient that the (pseudo) FPPs of its subsystems make a MVP have a full normal column/row rank which depends affinely on these (pseudo) FPPs. \textcolor{black}{Moreover, a matrix rank based necessary and sufficient condition is established for NDS structure identifiability under the following two situations. The first is that in each subsystem, no direct information transmission exists from an internal input to an internal output. The second is that in each subsystem, the TFM from its internal inputs to its external outputs can be expressed as the multiplication of a FNCR TFM and the TFM from its internal inputs to its internal outputs, while
the TFM from its external inputs to its internal outputs can be expressed as the multiplication of the TFM from its internal inputs to its internal outputs and a FNRR TFM.} This condition can be independently verified with each pair of two subsystems, and is therefore scalable for large scale NDSs.

From these results, it is conjectured that rather than the particular value of subsystem (pseudo) FPPs, it is the connections among subsystem states, internal/external inputs/outputs and (pseudo) FPPs that determine the structure identifiability of an NDS. That is, structure  identifiability of an NDS is possibly also a generic property, which is similar to its controllability and observability, as well as NDS identifiability with a prescribed structure. This is an interesting topic under investigations. In addition, further efforts are required to get a computationally scalable necessary and sufficient condition removing the assumptions like that on direct internal input-output information delivery. It is also interesting to develop a more appropriate metric measuring the distance of an NDS \textcolor{black}{to the set of NDSs with an unidentifiable structure}, taking into account both application significance and computation feasibility.

\vspace{0.25cm}
\hspace*{-0.45cm}{\rm\bf Acknowledgements.}
The author would like to thank Ms. K. L. Yin for various discussions and help in constructing the numerical example. Financial supports from NNSFC are also greatly appreciated.

\renewcommand{\theequation}{a.\arabic{equation}}
\setcounter{equation}{0}

\section*{Appendix: Proof of Some Technical Results}

\hspace*{-0.45cm}{\rm\bf Proof of Theorem \ref{theorem:1}:}
For each $i\in\{1,2,\cdots,N\}$, let $x(0,i)$ denote the initial value of the state vector of the $i$-th subsystem ${\rm\bf\Sigma}_{i}$. Take the Laplace transformation on both sides of Equation (\ref{eqn:1}) when $\delta(\cdot)$ is the derivative of a function with respect to time, and the $\cal Z$ transformation
when $\delta(\cdot)$ represents a forward time shift operation. Moreover, let $\star(\lambda,i)$ represent the associated signal after the transformation, in which $\star=x,u,v,y,z$. Then according to the properties of the Laplace/$\cal Z$ transformation, we have the following relations
\begin{eqnarray}
& &\!\!\!\! \left[\! \begin{array}{c}
\lambda x(\lambda,i)-x(0,i)\\
{z(\lambda,i)}\\
{{y}(\lambda,i)}
\end{array} \!\right] \nonumber\\
&=& \!\!\!\! \left[\! \begin{array}{ccc}
{A_{\rm\bf xx}{(i)}} & {A_{\rm\bf xv}{(i)}} & {B_{\rm\bf x}{(i)}}\\
{A_{\rm\bf zx}{(i)}} & {A_{\rm\bf zv}{(i)}} & {B_{\rm\bf z}{(i)}}\\
{C_{\rm\bf x}{(i)}} & {C_{\rm\bf v}{(i)}} & {D_{\rm\bf u}{(i)}}
\end{array} \!\right]\left[\! \begin{array}{c}
{x(\lambda,i)}\\
{v(\lambda,i)}\\
{u(\lambda,i)}
\end{array} \!\right]
\label{eqn:a1}
\end{eqnarray}

For each ${\rm\bf \#}={\it x}$, ${\it v}$ or ${\it z}$, define a vector $\#(\lambda)$ as $\#(\lambda)={\rm\bf col}\left\{\#(\lambda,i)|_{i=1}^{N}\right\}$. Moreover,
denote the vector ${\rm\bf col}\left\{x(0,i)|_{i=1}^{N}\right\}$ by $x(0)$. Furthermore,
define a matrix $D_{\rm\bf u}$  as $D_{\rm\bf u}\!\!=\!\!{\rm\bf diag}\!\left\{D_{\rm\bf u}(i)|_{i=1}^{N}\!\right\}$.  In addition, define matrices $A_{\rm\bf
*\#}$, $B_{\rm\bf *}$ and $C_{\rm\bf *}$ with ${\rm\bf *,\#}={\rm\bf x}$, ${\rm\bf y}$, ${\rm\bf v}$ or ${\rm\bf z}$ respectively as $A_{\rm\bf
*\#}=\!\!{\rm\bf diag}\!\left\{A_{\rm\bf
*\#}(i)|_{i=1}^{N}\!\right\}$, $B_{\rm\bf *}\!\!=\!\!{\rm\bf diag}\!\left\{B_{\rm\bf
*}(i)|_{i=1}^{N}\!\right\}$, $C_{\rm\bf *}\!=\!{\rm\bf diag}\!\left\{C_{\rm\bf
*}(i)|_{i=1}^{N}\!\right\}$. With these symbols, relations among all the transformed signals of all the subsystems in the NDS $\rm\bf\Sigma$, which is given by Equation (\ref{eqn:a1}), can be compactly represented by
\begin{equation}
\left[\! \begin{array}{c}
{\lambda x(\lambda)}-x(0)\\
{{{z}}(\lambda)}\\
{{y}(\lambda)}
\end{array} \!\right] = \left[\! \begin{array}{ccc}
{A_{\rm\bf xx}} & {A_{\rm\bf xv}} & {B_{\rm\bf x}}\\
{A_{\rm\bf zx}} & {A_{\rm\bf zv}} & {B_{\rm\bf z}}\\
{C_{\rm\bf x}} & {C_{\rm\bf v}} & {D_{\rm\bf u}}
\end{array} \!\right]\left[\! \begin{array}{c}
{x(\lambda)}\\
{{v}(\lambda)}\\
{u(\lambda)}
\end{array} \!\right]
\label{eqn:a2}
\end{equation}

From this equation, as well as the definitions of the TFMs $G_{\rm\bf zu}(\lambda)$, $G_{\rm\bf zv}(\lambda)$, $G_{\rm\bf yu}(\lambda)$ and $G_{\rm\bf yv}(\lambda)$,  direct algebraic manipulations show that
\begin{eqnarray}
\left[\! \begin{array}{c}
{{{z}}(\lambda)}\\
{{y}(\lambda)}
\end{array} \!\right] \!\!\!\! &=& \!\!\!\! \left[\! \begin{array}{cc}
{G_{\rm\bf zv}(\lambda)} & {G_{\rm\bf zu}(\lambda)}\\
{G_{\rm\bf yv}(\lambda)} & {G_{\rm\bf yu}(\lambda)}
\end{array} \!\right]\! \left[\! \begin{array}{c}
{v(\lambda)}\\
{u(\lambda)}
\end{array} \!\right]+ \nonumber\\
& & \!\!\!\!
\left[\! \begin{array}{c}
A_{\rm\bf zx} \\
C_{\rm\bf x}
\end{array} \!\right]\!
\left(\!\lambda I_{m_{\rm\bf x}} \!-\! A_{\rm\bf xx}\!\right)^{-1}x(0)
\label{eqn:a2-a}
\end{eqnarray}
in which $m_{\rm\bf x}={\sum_{k=1}^{N} m_{{\rm\bf x}k}}$.

On the other hand, from Equation (\ref{eqn:2}), we have that the following relation exists between the transformed internal input/output vectors of the NDS $\rm\bf\Sigma$,
\begin{equation}
v(\lambda)=\Phi z(\lambda)
\label{eqn:a3}
\end{equation}

Combining Equations (\ref{eqn:a2-a}) and (\ref{eqn:a3}) together, and recalling that the inverse of the TFM $I_{m_{\rm\bf z}}-G_{\rm\bf zv}(\lambda)\Phi$ is well defined when the NDS $\rm\bf\Sigma$ is well-posed, we immediately have that
\begin{equation}
y(\lambda,\Phi) = G(\lambda,\Phi)x(0) + H(\lambda,\Phi)u(\lambda)
\label{eqn:a4}
\end{equation}
Here, in order to clarify the dependence of NDS outputs on its SCM $\Phi$, the vector valued function $y(\lambda)$ is replaced by $y(\lambda,\Phi)$. In addition
\begin{eqnarray*}
& & \hspace*{-0.8cm} G(\lambda,\Phi) = \left\{ C_{\rm\bf x} + G_{\rm\bf yv}(\lambda)\Phi\left[ I_{m_{\rm\bf z}}-G_{\rm\bf zv}(\lambda)\Phi\right]^{-1}A_{\rm\bf zx}\right\}\times \\
& & \hspace*{5.5cm}
\left(\lambda I_{m_{\rm\bf x}}- A_{\rm\bf xx}\right)^{-1}
\end{eqnarray*}

Let $\Phi_{1}$ and $\Phi_{2}$ be two arbitrary different SCMs in the NDS $\rm\bf\Sigma$. Then with the same but arbitrary initial state vector $x(0)$ and the same also arbitrary inputs $u(\lambda)$, the difference between the associated outputs can be expressed as
\begin{eqnarray}
y(\lambda,\Phi_{1}) \!-\! y(\lambda,\Phi_{2})\!\!\!\!&=& \!\!\!\! \left[G(\lambda,\Phi_{1}) - G(\lambda,\Phi_{2})\right]x(0) +  \nonumber \\
& & \!\!\!\! \left[H(\lambda,\Phi_{1})-H(\lambda,\Phi_{2})\right] u(\lambda)
\label{eqn:a5}
\end{eqnarray}

Assume that there exist two different SCMs $\Phi_{1}$ and $\Phi_{2}$ for the NDS $\rm\bf\Sigma$, such that $H(\lambda,\Phi_{1})=H(\lambda,\Phi_{2})$ for every $\lambda\in {\cal C}$. The above equation implies that if all the initial states of the NDS $\rm\bf\Sigma$ are equal to zero, then for these two SCMs, we have that for each input series, the following equality holds,
\begin{equation}
y(\lambda,\Phi_{1}) = y(\lambda,\Phi_{2}), \hspace{0.5cm} \forall \lambda\in{\cal C}
\label{eqn:a6}
\end{equation}

Recall that both the Laplace transformation and the $\cal Z$-transformation are bijective mappings \cite{Kailath1980,zdg1996,zyl2018}. It is obvious that for these two SCMs $\Phi_{1}$ and $\Phi_{2}$, the associated outputs of the NDS $\rm\bf\Sigma(\Phi_{1})$ and the NDS $\rm\bf\Sigma(\Phi_{2})$ are always the same, no matter what input signals are used to stimulate them and how long the outputs are measured, provided that its initial states are all equal to zero. This implies that \textcolor{black}{the structure of this NDS $\rm\bf\Sigma$ is not identifiable.}

On the contrary, assume that for two arbitrary different SCMs $\Phi_{1}$ and $\Phi_{2}$ of the NDS $\rm\bf\Sigma$, there exist some $\lambda\in {\cal C}$ at which $H(\lambda,\Phi_{1}) \neq H(\lambda,\Phi_{2})$. Then according to Equation (\ref{eqn:a5}), if $\left[G(\lambda,\Phi_{1}) - G(\lambda,\Phi_{2})\right]x(0) \not\equiv 0$, then the input $u(t)|_{t=0}^{\infty}$ satisfying $u(\lambda)\equiv 0$, that is, the zero inputs, leads to $y(\lambda,\Phi_{1}) \not\equiv y(\lambda,\Phi_{2})$. In other words, the outputs of the NDS $\rm\bf\Sigma$ associated respectively with the SCMs $\Phi_{1}$ and $\Phi_{2}$ are different.

On the other hand, for an initial state vector $x(0)$ satisfying $\left[G(\lambda,\Phi_{1}) - G(\lambda,\Phi_{2})\right]x(0) =0$ for every $\lambda\in{\cal C}$, let $u(\lambda)= e_{j}$ in which $j$ is an element of the set consisting of the column indices of the TFM $H(\lambda,\Phi_{1}) -  H(\lambda,\Phi_{2})$ with which the corresponding column is not consistently equal to zero, while $e_{j}$ is the $j$-th standard basis of the ${m_{\rm\bf u}}$ dimensional Euclidean space ${\cal C}^{m_{\rm\bf u}}$ in which $m_{\rm\bf u}={\sum_{k=1}^{N} m_{{\rm\bf u}k}}$. From Equation (\ref{eqn:a5}), it is obvious that this input satisfies  $y(\lambda,\Phi_{1}) \not\equiv y(\lambda,\Phi_{2})$. That is, there exists at least one input time series, such that the outputs of the NDS $\rm\bf\Sigma(\Phi_{1})$ and the NDS $\rm\bf\Sigma(\Phi_{2})$ are not equal to each other at every time instant.

The above arguments mean that \textcolor{black}{if the aforementioned condition is satisfied, then the structure of the NDS $\rm\bf\Sigma$ is identifiable.} This completes the proof.   \hspace{\fill}$\Diamond$

\vspace{0.5cm}
\hspace*{-0.45cm}{\rm\bf Proof of Theorem \ref{theorem:2}:} Let $\Phi_{1}$ and $\Phi_{2}$ be two arbitrary SCMs satisfying the well-posedness assumption. Then both the TFM $I_{m_{\rm\bf z}} - G_{\rm\bf zv}(\lambda)\Phi_{1}$ and the TFM  $I_{m_{\rm\bf z}} - G_{\rm\bf zv}(\lambda)\Phi_{2}$ are of FNR. This implies that both the TFM $H(\lambda,\Phi_{1})$ and the TFM $H(\lambda,\Phi_{2})$ are well defined. From the definitions of the TFM $H(\lambda,\Phi)$, we have that
\begin{eqnarray}
& & \!\!\!\! H(\lambda,\Phi_{1}) - H(\lambda,\Phi_{2}) \nonumber \\
&=& \!\!\!\!  G_{\rm\bf yv}(\lambda)\left\{\!\Phi_{1}\left[I_{m_{\rm\bf z}} \!-\! G_{\rm\bf zv}(\lambda)\Phi_{1}\right]^{-1} \!-\! \right.\nonumber\\
& & \hspace*{3cm}\left.
\left[I_{m_{\rm\bf v}} \!-\! \Phi_{2}G_{\rm\bf zv}(\lambda)\right]^{-1}
\!\!\Phi_{2}\!\right\}G_{\rm\bf zu}(\lambda)  \nonumber\\
&=& \!\!\!\!  G_{\rm\bf yv}(\lambda)\left[I_{m_{\rm\bf v}} \!-\! \Phi_{2}G_{\rm\bf zv}(\lambda)\right]^{-1} \!\left(\Phi_{1} \!-\!\Phi_{2}\right) \times \nonumber\\
& & \hspace*{3cm} \left[I_{m_{\rm\bf z}} \!-\! G_{\rm\bf zv}(\lambda)\Phi_{1}\right]^{-1}\!\!G_{\rm\bf zu}(\lambda)  \nonumber\\
&=& \!\!\!\!  G_{\rm\bf yv}(\lambda) \Delta(\lambda) G_{\rm\bf zu}(\lambda)
\label{eqn:a7}
\end{eqnarray}
in which
\begin{displaymath}
\Delta(\lambda)=\left[I_{m_{\rm\bf v}} \!-\! \Phi_{2}G_{\rm\bf zv}(\lambda)\right]^{-1} \!\left(\Phi_{1} \!-\!\Phi_{2}\right) \left[I_{m_{\rm\bf z}} \!-\! G_{\rm\bf zv}(\lambda)\Phi_{1}\right]^{-1}
\end{displaymath}

Note that
\begin{displaymath}
{\rm\bf det}\left\{I_{m_{\rm\bf v}} \!-\! \Phi_{2}G_{\rm\bf zv}(\lambda)\right\} \!=\!{\rm\bf det}\left\{I_{m_{\rm\bf z}} \!-\! G_{\rm\bf zv}(\lambda)\Phi_{2}\right\}
\end{displaymath}
This means that the TFM $I_{m_{\rm\bf v}} \!-\! \Phi_{2}G_{\rm\bf zv}(\lambda)$ is also of FNR and invertible for almost each  $\lambda\in{\cal C}$. These imply that if $\Phi_{1}=\Phi_{2}$ then $\Delta(\lambda)=0$ for all the $\lambda\in{\cal C}$.

On the contrary, assume that $\Delta(\lambda)=0$ for almost all the $\lambda\in{\cal C}$. As both the TFM $I_{m_{\rm\bf v}} \!-\! \Phi_{2}G_{\rm\bf zv}(\lambda)$ and the TFM $I_{m_{\rm\bf z}} \!-\! G_{\rm\bf zv}(\lambda)\Phi_{1}$ are of FNR, it can be proven using arguments similar to those between the following Equations (\ref{eqn:a8}) and (\ref{eqn:a13}), that there certainly exists at least one $\lambda_{0}\in{\cal C}$, such that $\Delta(\lambda_{0})=0$, while $I_{m_{\rm\bf v}} \!-\! \Phi_{2}G_{\rm\bf zv}(\lambda_{0})$ and $I_{m_{\rm\bf z}} \!-\! G_{\rm\bf zv}(\lambda_{0})\Phi_{1}$ are invertible. From the definition of the TFM $\Delta(\lambda)$, this means that $\Phi_{1}=\Phi_{2}$.

The above arguments mean that $\Phi_{1}=\Phi_{2}$ if and only if $\Delta(\lambda)=0$ for all the $\lambda\in{\cal C}$.

On the other hand, according to the Smith-McMillan form of a TFM, it can be declared that there exist a nonnegative integer $\overline{m}_{\rm\bf z}$ not greater than $m_{\rm\bf z}$, an $m_{\rm\bf z}\times m_{\rm\bf z}$ dimensional unimodular matrix $U_{\rm\bf zu}(\lambda)$, an $m_{\rm\bf u}\times m_{\rm\bf u}$ dimensional unimodular matrix $V_{\rm\bf zu}(\lambda)$, as well as nonzero and real coefficient polynomials $\alpha^{[i]}_{\rm\bf zu}(\lambda)|_{i=1}^{\overline{m}_{\rm\bf z}}$ and  $\beta^{[i]}_{\rm\bf zu}(\lambda)|_{i=1}^{\overline{m}_{\rm\bf z}}$ with all of them having a finite degree, such that
\begin{equation}
G_{\rm\bf zu}(\lambda)=U_{\rm\bf zu}(\lambda)
\!\left[\!\begin{array}{cc}
{\rm\bf diag}\left\{\left.\frac{\alpha^{[i]}_{\rm\bf zu}(\lambda)}{\beta^{[i]}_{\rm\bf zu}(\lambda)}\right|_{i=1}^{\overline{m}_{\rm\bf z}}\right\} & 0 \\
0 & 0 \end{array}\!\right]\!
V_{\rm\bf zu}(\lambda)
\label{eqn:a8}
\end{equation}
Here, the dimensions of the zero matrices are in general different. They are not clearly indicated for brevity.

Divide the unimodular matrix $U_{\rm\bf zu}(\lambda)$ as $U_{\rm\bf zu}(\lambda)=\left[U_{\rm\bf zu}^{[1]}(\lambda)\; U_{\rm\bf zu}^{[2]}(\lambda)\right]$ with $U_{\rm\bf zu}^{[1]}(\lambda)$ having $\overline{m}_{\rm\bf z}$ columns. Then from Equation (\ref{eqn:a8}), we have that
\begin{equation}
G_{\rm\bf zu}(\lambda)=\left[ U_{\rm\bf zu}^{[1]}(\lambda)
{\rm\bf diag}\left\{\left.\frac{\alpha^{[i]}_{\rm\bf zu}(\lambda)}{\beta^{[i]}_{\rm\bf zu}(\lambda)}\right|_{i=1}^{\overline{m}_{\rm\bf z}}\right\} \;\;  0 \right]\!
V_{\rm\bf zu}(\lambda)
\label{eqn:a9}
\end{equation}

As $U_{\rm\bf zu}(\lambda)$ is an unimodular matrix, there exists another unimodular matrix $U_{\rm\bf zu}^{[iv]}(\lambda)$, such that
\begin{equation}
U_{\rm\bf zu}^{[iv]}(\lambda)U_{\rm\bf zu}(\lambda)=I_{m_{\rm\bf z}}
\label{eqn:a10}
\end{equation}
Partition the unimodular matrix $U_{\rm\bf zu}^{[iv]}(\lambda)$ as $U_{\rm\bf zu}^{[iv]}(\lambda)={\rm\bf col}\left\{U_{\rm\bf zu,1}^{[iv]}(\lambda),\; U_{\rm\bf zu,2}^{[iv]}(\lambda)\right\}$ with $U_{\rm\bf zu,1}^{[iv]}(\lambda)$ having $\overline{m}_{\rm\bf z}$ rows. It can then be declared from Equation (\ref{eqn:a10}) that
\begin{equation}
U_{\rm\bf zu,2}^{[iv]}(\lambda)U_{\rm\bf zu}^{[1]}(\lambda)\equiv 0
\label{eqn:a11}
\end{equation}

Construct a polynomial vector $\zeta(\lambda)$ as
\begin{equation}
\zeta(\lambda)=\xi(\lambda)U_{\rm\bf zu,2}^{[iv]}(\lambda)
\label{eqn:a12}
\end{equation}
in which $\xi(\lambda)$ is an arbitrary $m_{\rm\bf z}-\overline{m}_{\rm\bf z}$ dimensional row polynomial vector with real coefficients that does not make the associated polynomial vector $\zeta(\lambda)$ being equal to zero at each $\lambda\in{\cal C}$. The existence of this polynomial vector is guaranteed by the fact that the MVP $U_{\rm\bf zu}^{[iv]}(\lambda)$ is unimodular, which means that the sub-MVP $U_{\rm\bf zu,2}^{[iv]}(\lambda)$ is of FRR at each complex $\lambda$.
Substitute this $\zeta(\lambda)$ into Equation (\ref{eqn:a9}). It is immediate from Equation (\ref{eqn:a11}) that
\begin{eqnarray}
& & \!\!\!\! \zeta(\lambda) G_{\rm\bf zu}(\lambda) \nonumber \\
&=& \!\!\!\!  \xi(\lambda)U_{\rm\bf zu,2}^{[iv]}(\lambda)\!
\left[ U_{\rm\bf zu}^{[1]}(\lambda)
{\rm\bf diag}\left\{\left.\frac{\alpha^{[i]}_{\rm\bf zu}(\lambda)}{\beta^{[i]}_{\rm\bf zu}(\lambda)}\right|_{i=1}^{\overline{m}_{\rm\bf z}}\right\} \;\;  0 \right]\!
V_{\rm\bf zu}(\lambda)
\nonumber \\
&\equiv& 0
\label{eqn:a13}
\end{eqnarray}

The above arguments show that if the integer $\overline{m}_{\rm\bf z}$ is smaller than $m_{\rm\bf z}$, then the TFM $G_{\rm\bf zu}(\lambda)$ is row rank deficient at every $\lambda\in{\cal C}$, and is therefore not of FNRR.

Assume now that for each $i\in \{1,2,\cdots,N\}$, the TFM $G_{\rm\bf zu}(\lambda,i)$ is of FNRR, while the TFM $G_{\rm\bf yv}(\lambda,i)$ is of FNCR. From the block diagonal structure of the  TFMs $G_{\rm\bf zu}(\lambda)$ and $G_{\rm\bf yv}(\lambda)$, it can be directly declared that the TFM $G_{\rm\bf zu}(\lambda)$ is of FNRR, while the TFM $G_{\rm\bf yv}(\lambda)$ is of FNCR.

From these observations and Equation (\ref{eqn:a13}), it is clear that there exist an $m_{\rm\bf z}\times m_{\rm\bf z}$ dimensional unimodular matrix $U_{\rm\bf zu}(\lambda)$, an $m_{\rm\bf u}\times m_{\rm\bf u}$ dimensional unimodular matrix $V_{\rm\bf zu}(\lambda)$, as well as nonzero and real coefficient polynomials $\alpha^{[i]}_{\rm\bf zu}(\lambda)|_{i=1}^{m_{\rm\bf z}}$ and  $\beta^{[i]}_{\rm\bf zu}(\lambda)|_{i=1}^{m_{\rm\bf z}}$ with all of them having a finite degree, such that
\begin{equation}
G_{\rm\bf zu}(\lambda)=U_{\rm\bf zu}(\lambda)
\!\left[\!\begin{array}{cc}
{\rm\bf diag}\left\{\left.\frac{\alpha^{[i]}_{\rm\bf zu}(\lambda)}{\beta^{[i]}_{\rm\bf zu}(\lambda)}\right|_{i=1}^{m_{\rm\bf z}}\right\} & 0 \end{array}\!\right]\!
V_{\rm\bf zu}(\lambda)
\label{eqn:a8-a}
\end{equation}
In addition, noting that a TFM is of FNCR if and only if its transpose is of FNRR. This implies that there also exist an $m_{\rm\bf y}\times m_{\rm\bf y}$ dimensional unimodular matrix $U_{\rm\bf yv}(\lambda)$, an $m_{\rm\bf v}\times m_{\rm\bf v}$ dimensional unimodular matrix $V_{\rm\bf yv}(\lambda)$, as well as nonzero and real coefficient polynomials $\alpha^{[i]}_{\rm\bf yv}(\lambda)|_{i=1}^{m_{\rm\bf v}}$ and  $\beta^{[i]}_{\rm\bf yv}(\lambda)|_{i=1}^{m_{\rm\bf v}}$ with all of them having a finite degree, such that
\begin{equation}
G_{\rm\bf yv}(\lambda)=U_{\rm\bf yv}(\lambda)
\!\left[\!\begin{array}{c}
{\rm\bf diag}\left\{\left.\frac{\alpha^{[i]}_{\rm\bf yv}(\lambda)}{\beta^{[i]}_{\rm\bf yv}(\lambda)}\right|_{i=1}^{m_{\rm\bf v}}\right\} \\ 0 \end{array}\!\right]\!
V_{\rm\bf yv}(\lambda)
\label{eqn:a8-b}
\end{equation}

Equations (\ref{eqn:a8-a}) and (\ref{eqn:a8-b}) mean that the TFM $G_{\rm\bf zu}(\lambda)$ is right invertible for almost every $\lambda\in{\cal C}$, while the TFM $G_{\rm\bf yv}(\lambda)$ is left invertible for almost every $\lambda\in{\cal C}$.

More precisely, define sets ${\rm\bf\Lambda}_{\rm\bf zu}$ and ${\rm\bf\Lambda}_{\rm\bf yv}$ respectively as
\begin{eqnarray*}
& & {\rm\bf\Lambda}_{\rm\bf zu}=\bigcup_{i=1}^{m_{\rm\bf z}}\left\{\lambda\left|
\alpha^{[i]}_{\rm\bf zu}(\lambda)=0,\hspace{0.2cm}\lambda\in{\cal C}\right.\right\} \\
& & {\rm\bf\Lambda}_{\rm\bf yv}=\bigcup_{i=1}^{m_{\rm\bf v}}\left\{\lambda\left|
\alpha^{[i]}_{\rm\bf yv}(\lambda)=0,\hspace{0.2cm}\lambda\in{\cal C}\right.\right\}
\end{eqnarray*}
As both the polynomials $\alpha^{[i]}_{\rm\bf zu}(\lambda)|_{i=1}^{m_{\rm\bf z}}$ and the polynomials $\alpha^{[i]}_{\rm\bf yv}(\lambda)|_{i=1}^{m_{\rm\bf v}}$ are of finite degree, it is obvious that each of these two sets has only finite elements. On the other hand, from Equations (\ref{eqn:a8-a}) and (\ref{eqn:a8-b}), it is clear that the TFM $G_{\rm\bf zu}(\lambda,i)$ is not of FRR only when $\lambda\in {\rm\bf\Lambda}_{\rm\bf zu}$, while the TFM $G_{\rm\bf yv}(\lambda,i)$ is not of FCR only when $\lambda\in {\rm\bf\Lambda}_{\rm\bf yv}$. Therefore for every $\lambda\in {\cal C}\left/\left\{{\rm\bf\Lambda}_{\rm\bf zu}\bigcup
{\rm\bf\Lambda}_{\rm\bf yv}\right\}\right.$, the TFM $G_{\rm\bf zu}(\lambda)$ is of FRR and the TFM $G_{\rm\bf yv}(\lambda)$ is of FCR, and hence are respectively right and left invertible.

Combining these observations with Equation (\ref{eqn:a7}), it can be declared that if $H(\lambda,\Phi_{1}) = H(\lambda,\Phi_{2})$ for all the $\lambda\in{\cal C}$, then for each $\lambda\in {\cal C}\left/\left\{{\rm\bf\Lambda}_{\rm\bf zu}\bigcup
{\rm\bf\Lambda}_{\rm\bf yv}\right\}\right.$, $\Delta(\lambda)=0$. This further implies that $\Delta(\lambda)=0$ for every $\lambda\in{\cal C}$. Hence, $\Phi_{1}=\Phi_{2}$. That is, \textcolor{black}{the structure of the NDS $\rm\bf\Sigma$ is identifiable.}

This completes the proof. \hspace{\fill}$\Diamond$

\vspace{0.5cm}
\hspace*{-0.45cm}{\rm\bf Proof of Theorem \ref{theorem:5}:}  Let $\Phi_{1}$ and $\Phi_{2}$ be two arbitrary SCMs. When the TFM $G_{\rm\bf zv}(\lambda,i)$ is \textcolor{black}{identically} equal to zero for each $i=1,2,\cdots,N$, the TFM $G_{\rm\bf zv}(\lambda)$ also has this property. Hence, both the TFM $I_{m_{\rm\bf z}} \!-\! G_{\rm\bf zv}(\lambda)\Phi_{1}$ and the TFM  $I_{m_{\rm\bf z}} \!-\! G_{\rm\bf zv}(\lambda)\Phi_{2}$ are in fact the identity matrix $I_{m_{\rm\bf z}}$ and are therefore always of FNR. That is, the associated NDSs ${\rm\bf\Sigma}(\Phi_{1})$ and ${\rm\bf\Sigma}(\Phi_{2})$ are always well-posed.

On the other hand, under the situation that $G_{\rm\bf zv}(\lambda)\equiv 0$, it is obvious from Equation (\ref{eqn:a7}) that
\begin{equation}
H(\lambda,\Phi_{1}) - H(\lambda,\Phi_{2}) =
G_{\rm\bf yv}(\lambda) (\Phi_{1}-\Phi_{2}) G_{\rm\bf zu}(\lambda)
\label{eqn:a40}
\end{equation}
Partition the SCMs $\Phi_{1}$ and $\Phi_{2}$ respectively as
\begin{equation}
\Phi_{1}=\left[\Phi_{1}(i,j)|_{i,j=1}^{N}\right], \hspace{0.25cm}
\Phi_{2}=\left[\Phi_{2}(i,j)|_{i,j=1}^{N}\right]
\label{eqn:a41}
\end{equation}
in which $\Phi_{1}(i,j)$ and $\Phi_{2}(i,j)$ are $m_{{\rm\bf v}i}\times {m_{{\rm\bf z}j}}$ dimensional real submatrix. Moreover, denote $\Phi_{1}(i,j)-\Phi_{2}(i,j)$ with $i,j=1,2,\cdots,N$, by $\Delta(i,j)$ for brevity. Then from the consistent block diagonal structure of the TFMs $G_{\rm\bf yv}(\lambda)$ and $G_{\rm\bf zu}(\lambda)$, it is immediate that $H(\lambda,\Phi_{1}) - H(\lambda,\Phi_{2})\equiv 0$ if and only if for every $i,j=1,2,\cdots,N$,
\begin{equation}
G_{\rm\bf yv}(\lambda,i) \Delta(i,j) G_{\rm\bf zu}(\lambda,j) \equiv 0
\label{eqn:a42}
\end{equation}

Substitute Equations (\ref{eqn:29}) and (\ref{eqn:30}) into the above equation. Noting that both the MVPs $U_{\rm\bf yv}(\lambda,i)$ and $V_{\rm\bf zu}(\lambda,i)$ are unimodular, as well as that the polynomials $\left.\alpha^{[j]}_{\rm\bf yv}(\lambda,i)\right|_{j=1}^{\overline{m}_{{\rm\bf v}i}}$ and $\left.\alpha^{[j]}_{\rm\bf zu}(\lambda,i)\right|_{j=1}^{\overline{m}_{{\rm\bf z}i}}$ are not identically equal to zero and have a finite degree, it can be straightforwardly shown that Equation (\ref{eqn:a42}) is satisfied, if and only if
\begin{equation}
V^{[1]}_{\rm\bf yv}(\lambda,i)\Delta(i,j)U^{[1]}_{\rm\bf zu}(\lambda,j)\equiv 0
\label{eqn:a43}
\end{equation}

In addition, from Equation (\ref{eqn:31}), we have that
\begin{eqnarray}
& & \!\!\!\! V^{[1]}_{\rm\bf yv}(\lambda,i)\Delta(i,j)U^{[1]}_{\rm\bf zu}(\lambda,j) \nonumber \\
&=& \!\!\!\!
\left(\!\sum_{p=0}^{d^{[vi1]}_{\rm\bf yv}}
\! V_{\rm\bf yv}^{[1]}(i,p)\lambda^{p}\!\right) \Delta(i,j) \left(\!
\sum_{q=0}^{d^{[uj1]}_{\rm\bf zu}}\!
U_{\rm\bf zu}^{[1]}(j,q)\lambda^{q}\!\right) \nonumber\\
&=& \!\!\!\!
\sum_{p=0}^{d^{[vi1]}_{\rm\bf yv}}\sum_{q=0}^{d^{[uj1]}_{\rm\bf zu}}\!
 V_{\rm\bf yv}^{[1]}(i,p)\Delta(i,j)
U_{\rm\bf zu}^{[1]}(j,q)\lambda^{p+q} \nonumber\\
&=& \!\!\!\!
\sum_{k=0}^{d^{[vi1]}_{\rm\bf yv}+d^{[uj1]}_{\rm\bf zu}}\!\!\!\!\left( \sum_{s=\max\{0,k-d^{[uj1]}_{\rm\bf zu}\}}^{\min\{k,d^{[uj1]}_{\rm\bf zu}\}}\!
\!\!\!\!\!\!  V_{\rm\bf yv}^{[1]}(i,k\!-\!s)\Delta(i,j)
U_{\rm\bf zu}^{[1]}(j,s)\!\!\right)\!\lambda^{k}  \nonumber\\
& &
\label{eqn:a44}
\end{eqnarray}
Therefore, Equation (\ref{eqn:a43}) is satisfied, if and only if for each $k=0,1,\cdots, d^{[vi1]}_{\rm\bf yv}+d^{[uj1]}_{\rm\bf zu}$,
\begin{equation}
\sum_{s=\max\{0,k-d^{[uj1]}_{\rm\bf zu}\}}^{\min\{k,d^{[uj1]}_{\rm\bf zu}\}}\!
\!\!\!\! V_{\rm\bf yv}^{[1]}(i,k\!-\!s)\Delta(i,j)
U_{\rm\bf zu}^{[1]}(j,s) \!=\! 0
\label{eqn:a45}
\end{equation}
which is equivalent to
\begin{eqnarray}
& & \!\!\!\!{\rm\bf vec}\!\left(\!\sum_{s=\max\{0,k-d^{[uj1]}_{\rm\bf zu}\}}^{\min\{k,d^{[uj1]}_{\rm\bf zu}\}}\!
\!\!\!\! V_{\rm\bf yv}^{[1]}(i,s)\Delta(i,j)
U_{\rm\bf zu}^{[1]}(j,k\!-\!s
)\!\! \right) \nonumber\\
&=& \!\!\!\! \left[
\sum_{s=\max\{0,k-d^{[uj1]}_{\rm\bf zu}\}}^{\min\{k,d^{[uj1]}_{\rm\bf zu}\}}\!
\!\!\!\! U_{\rm\bf zu}^{[1]T}(j,k\!-\!s)\otimes V_{\rm\bf yv}^{[1]}(i,s)\!\right]\! {\rm\bf vec}\!\left(\!\Delta(i,j)\!\right) \nonumber\\
&=& \!\!\!\! 0
\label{eqn:a46}
\end{eqnarray}

Assume now that the matrix $\Xi_{\rm\bf zu}^{\rm\bf yv}(i,j)$ is of FCR. Then Equations (\ref{eqn:a43})-(\ref{eqn:a46}) mean that Equation (\ref{eqn:a42}) has a unique solution $\Delta(i,j)=0$, and vice versa. It can therefore be declared from Theorem \ref{theorem:1} that the condition that the matrix $\Xi_{\rm\bf zu}^{\rm\bf yv}(i,j)$ is of FCR for each $i,j=1,2,\cdots,N$, is both necessary and sufficient for \textcolor{black}{the structure of the NDS $\rm\bf\Sigma$ being identifiable.}

This completes the proof. \hspace{\fill}$\Diamond$

\vspace{0.5cm}
\hspace*{-0.45cm}{\rm\bf Proof of Theorem \ref{theorem:3}:} From Equation (\ref{eqn:14}), we have that
\begin{eqnarray}
& & \hspace*{-1cm} G_{\rm\bf yv}(\lambda,i) \! = \!
H_{\rm\bf yv}(\lambda,i) \! +\!
H_{\rm\bf yr}(\lambda,i) P(i) \times \nonumber\\
& & \hspace*{1cm} \left[I_{m_{{\rm\bf w}i}} \!-\! H_{\rm\bf wr}(\lambda,i)P(i)\right]^{-1} \!\!H_{\rm\bf wv}(\lambda,i)
\label{eqn:a14}
\end{eqnarray}

For a particular $\lambda\in {\cal C}$, assume that there is a vector $\alpha$ satisfying $G_{\rm\bf yv}(\lambda,i)\alpha=0$. Define a vector $\beta$ as \begin{equation}
\beta = P(i)\left[I_{m_{{\rm\bf w}i}} \!-\! H_{\rm\bf wr}(\lambda,i)P(i)\right]^{-1} \!\!H_{\rm\bf wv}(\lambda,i)\alpha
\label{eqn:a21}
\end{equation}
Obviously, the vector $\beta$ can also be expressed as
\begin{equation}
\beta = \left[I_{m_{{\rm\bf p}i}} \!-\! P(i)H_{\rm\bf wr}(\lambda,i)\right]^{-1} \!\!P(i)H_{\rm\bf wv}(\lambda,i)\alpha
\label{eqn:a22}
\end{equation}
Hence, the vectors $\alpha$ and $\beta$ satisfy
\begin{equation}
\left[\begin{array}{cc}
H_{\rm\bf yv}(\lambda,i)  &  H_{\rm\bf yr}(\lambda,i)  \\
P(i)H_{\rm\bf wv}(\lambda,i) & P(i)H_{\rm\bf wr}(\lambda,i) \!-\! I_{m_{{\rm\bf p}i}}  \end{array}\right]\!\!
\left[\begin{array}{c} \alpha  \\  \beta \end{array}\right] = 0
\label{eqn:a15}
\end{equation}

On the other hand, from Equation (\ref{eqn:12}), it can be straightforwardly proven that
\begin{eqnarray}
& &\hspace*{-0.8cm}
\left[\begin{array}{cc}
H_{\rm\bf yv}(\lambda,i)  &  H_{\rm\bf yr}(\lambda,i)  \\
P(i)H_{\rm\bf wv}(\lambda,i) & P(i)H_{\rm\bf wr}(\lambda,i) \!-\! I_{m_{{\rm\bf p}i}}  \end{array}\right] \nonumber\\
& & \hspace*{-1.0cm}=\!\!
\left[\!\!\begin{array}{cc}
0  &  0  \\
0 & -\! I_{m_{{\rm\bf p}i}}  \end{array}\!\!\!\right]
\!+\! \left[\!\!\begin{array}{cc}
I &  0  \\
0 & P(i)  \end{array}\!\!\right]\!\!\left\{\!
\left[\begin{array}{cc} {C^{[0]}_{\rm\bf v}(i)} & H_{\rm\bf y}(i) \\ F_{\rm\bf v}(i) & G(i)
\end{array}  \!\!\right]\!+ \right.\nonumber  \\
& & \hspace*{-0.8cm}
\left. \left[\!\! \begin{array}{c}
{C^{[0]}_{\rm\bf x}(i)}  \\
F_{\rm\bf x}(i)  \end{array}\!\! \right]\!\!\left(\lambda I_{m_{{\rm\bf x}i}} \!-\! {A^{[0]}_{\rm\bf xx}(i)}\right)^{-1}\!\!
\left[\!\! \begin{array}{cc}
{A^{[0]}_{\rm\bf xv}(i)} & H_{\rm\bf x}(i) \end{array}\!\! \right] \!\!\right\}
\label{eqn:a16}
\end{eqnarray}
in which the zero matrices in general have different dimensions.

Define a vector $\xi$ as
\begin{equation}
\xi = \left(\lambda I_{m_{{\rm\bf x}i}} \!-\! {A^{[0]}_{\rm\bf xx}(i)}\right)^{-1}\!\!
\left[\!\! \begin{array}{cc}
{A^{[0]}_{\rm\bf xv}(i)} & H_{\rm\bf x}(i) \end{array}\!\! \right]
\left[\begin{array}{c} \alpha  \\  \beta \end{array}\right]
\label{eqn:a17}
\end{equation}
Then we have that
\begin{equation}
\left[\!\! \begin{array}{ccc}
\lambda I_{m_{{\rm\bf x}i}} \!-\! A^{[0]}_{\rm\bf xx}(i) & -{A^{[0]}_{\rm\bf xv}(i)} & -H_{\rm\bf x}(i) \end{array}\!\! \right]\!\!
\left[\begin{array}{c} \xi \\ \alpha  \\  \beta \end{array}\!\!\right] = 0
\label{eqn:a18}
\end{equation}

Moreover, from Equations (\ref{eqn:a15}) and (\ref{eqn:a16}), as well as the definition of the vector $\xi$, direct matrix manipulations show that
\begin{equation}
\hspace*{-0.0cm}
\left[\!\! \begin{array}{ccc}
C^{[0]}_{\rm\bf x}(i)  &  C^{[0]}_{\rm\bf v}(i) &  H_{\rm\bf y}(i)  \\
P(i)F_{\rm\bf x}(i) & P(i)F_{\rm\bf v}(i)  & P(i)G(i)- I_{m_{{\rm\bf p}i}}
\end{array}\!\! \right] \!\!
\left[\!\! \begin{array}{c}
\xi \\
\alpha \\
\beta \end{array}\!\! \right] = 0
\label{eqn:a19}
\end{equation}

Combining Equations (\ref{eqn:a18}) and (\ref{eqn:a19}) together, the definition of the matrix pencil $M(\lambda,i)$ leads immediately to the following equality,
\begin{equation}
M(\lambda,i) {\rm\bf col}\!\left\{\xi,\; \alpha, \;\beta \right\} = 0
\label{eqn:a20}
\end{equation}

Assume now that the TFM $G_{\rm\bf yv}(\lambda,i)$ is not of FNCR. Then for an arbitrary $\lambda\in {\cal C}$, there exists an nonzero vector $\alpha$ satisfying $G_{\rm\bf yv}(\lambda,i)\alpha=0$. The above arguments show that under such a situation, the corresponding vector ${\rm\bf col}\!\left\{\xi,\; \alpha, \;\beta \right\}$ with its sub-vectors $\beta$ and $\xi$ being defined respectively by Equations (\ref{eqn:a21}) and (\ref{eqn:a17}), is also nonzero and satisfies Equation (\ref{eqn:a20}). This means that the matrix pencil $M(\lambda,i)$ is not of FNCR, also.

On the contrary, assume that the matrix pencil $M(\lambda,i)$ is not of FNCR. Then for each $\lambda\in {\cal C}$, there exists at least one nonzero vector $\zeta$ such that $M(\lambda,i)\zeta=0$. Partition this vector $\zeta$ as
\begin{equation}
\zeta = {\rm\bf col}\!\left\{\xi,\; \alpha, \;\beta \right\}
\label{eqn:a23}
\end{equation}
with the sub-vectors $\xi$, $\alpha$ and $\beta$ having respectively $m_{{\rm\bf x}i}$, $m_{{\rm\bf v}i}$ and $m_{{\rm\bf p}i}$ elements. On the basis of Equation (\ref{eqn:a20}), direct algebraic manipulations show that the sub-vector $\alpha$ must not be a  zero vector and satisfies $G_{\rm\bf yv}(\lambda,i)\alpha=0$. Hence, the TFM $G_{\rm\bf yv}(\lambda,i)$ is also not of FNCR.

This completes the proof.   \hspace{\fill}$\Diamond$

\vspace{0.5cm}
\hspace*{-0.45cm}{\rm\bf Proof of Corollary \ref{cor:1}:}

Substitute the KCF of Equation (\ref{eqn:20}) into Equation (\ref{eqn:19}), the following equality is obtained,
\begin{equation}
\overline{M}(\lambda,i)\!= \! {\rm\bf diag}\!\left\{ U(i),\;I_{{\rm\bf p}i} \right\}
\hat{M}(\lambda,i)V(i)
\label{eqn:a24}
\end{equation}
in which
\begin{equation}
\hspace*{-0.6cm}\hat{M}(\lambda,i)\!= \!
\left[\!\!\!\!\!\begin{array}{c}
K(\lambda,i) \\
\begin{array}{l}P(i)\left[F_{\rm\bf x}(i)N_{\rm\bf x}(i) + F_{\rm\bf v}(i)N_{\rm\bf v}(i) + \right. \\
\hspace*{0.5cm}\left.
G(i)N_{\rm\bf w}(i) \right]V^{-1}(i) \!-\!
N_{\rm\bf w}(i)V^{-1}(i)\end{array}  \end{array}\!\!\!\!\!\right]
\label{eqn:a25}
\end{equation}

Note that both the matrix $U(i)$ and the matrix $V(i)$ are invertible and independent of the complex variable $\lambda$. It is obvious that the matrix pencil $\overline{M}(\lambda,i)$ is of FNCR, if and only if the matrix pencil $\hat{M}(\lambda,i)$ is. As the matrix pencil $\tilde{M}(\lambda,i)$ is in fact the sub-matrix of the matrix pencil $\hat{M}(\lambda,i)$ constituted from its first ${\rm\bf m}(i)$ columns, this means that the matrix pencil $\overline{M}(\lambda,i)$ is of FNCR, only if the matrix pencil $\tilde{M}(\lambda,i)$ is.

On the contrary, assume that the matrix pencil $\tilde{M}(\lambda,i)$ is of FNCR.
Then there exists at least one $\lambda_{0}\in{\cal C}$, such that for an arbitrary ${\rm\bf m}(i)$ dimensional nonzero complex vector $\zeta$, the matrix $\tilde{M}(\lambda_{0},i)$ satisfies $\tilde{M}(\lambda_{0},i)\zeta\neq 0$.

On the other hand, according to the Smith form of a MVP,  there exist a nonnegative integer $\overline{\rm\bf m}(i)$, an $\left(\sum_{j=1}^{\zeta^{[i]}_{{\rm\bf L}}}{\xi^{[i]}_{\rm\bf L}(j)}+ m_{{\rm\bf p}i}\right)\times \left(\sum_{j=1}^{\zeta^{[i]}_{{\rm\bf L}}}{\xi^{[i]}_{\rm\bf L}(j)}+ m_{{\rm\bf p}i}\right)$ dimensional unimodular matrix $\tilde{U}(\lambda,i)$, an ${\rm\bf m}(i)\times {\rm\bf m}(i)$ dimensional unimodular matrix $\tilde{V}(\lambda,i)$, as well as nonzero and real coefficient polynomials $\tilde{\alpha}^{[j]}(\lambda)|_{j=1}^{\overline{\rm\bf m}(i)}$ with a finite degree, such that
\begin{equation}
\tilde{M}(\lambda,i)=\tilde{U}(\lambda,i)
\!\left[\!\begin{array}{cc}
{\rm\bf diag}\left\{\left.\tilde{\alpha}^{[j]}(\lambda)\right|_{j=1}^{\overline{\rm\bf m}(i)}\right\} & 0 \\
0 & 0 \end{array}\!\right]\!
\tilde{V}(\lambda,i)
\label{eqn:a27}
\end{equation}
in which the zero matrices may not have the same dimension.

Assume that $\overline{\rm\bf m}(i)<{\rm\bf m}(i)$.
Divide the unimodular matrix $\tilde{V}(\lambda,i)$ as $\tilde{V}(\lambda,i)={\rm\bf col}\!\left\{\tilde{V}_{1}(\lambda,i),\; \tilde{V}_{2}(\lambda,i)\right\}$ with $\tilde{V}_{1}(\lambda,i)$ having $\overline{\rm\bf m}(i)$ rows. Then the sub-MVP $\tilde{V}_{2}(\lambda,i)$ is not empty. Moreover, from Equation (\ref{eqn:a27}), we have that
\begin{equation}
\tilde{M}(\lambda,i)=\tilde{U}(\lambda,i)
\!\left[\!\begin{array}{c}
{\rm\bf diag}\left\{\left.\tilde{\alpha}^{[j]}(\lambda)\right|_{j=1}^{\overline{\rm\bf m}(i)}\right\}  \\
0  \end{array}\!\right]\!\tilde{V}_{1}(\lambda,i)
\label{eqn:a28}
\end{equation}

Note that $\tilde{V}(\lambda,i)$ is an unimodular matrix. There exists another unimodular matrix $\tilde{V}^{[iv]}(\lambda,i)$ satisfying
\begin{equation}
\tilde{V}(\lambda,i)\tilde{V}^{[iv]}(\lambda,i) \equiv I_{{\rm\bf m}(i)}
\label{eqn:a29}
\end{equation}
Partition the unimodular matrix $\tilde{V}^{[iv]}(\lambda,i)$ as $\tilde{V}^{[iv]}(\lambda,i)=\left[\tilde{V}_{1}^{[iv]}(\lambda,i)\;\; \tilde{V}_{2}^{[iv]}(\lambda,i)\right]$ with $\tilde{V}_{2}^{[iv]}(\lambda,i)$ having ${\rm\bf m}(i)-\overline{\rm\bf m}(i)$ columns. It can then be declared from Equation (\ref{eqn:a29}) that the sub-MVP $\tilde{V}_{2}^{[iv]}(\lambda,i)$ is of FCR at every $\lambda\in {\cal C}$, and
\begin{equation}
\tilde{V}_{1}(\lambda,i)\tilde{V}_{2}^{[iv]}(\lambda,i)\equiv 0
\label{eqn:a30}
\end{equation}

Combining Equations (\ref{eqn:a27})-(\ref{eqn:a30}) together, we have that for an arbitrary $\lambda\in{\cal C}$ and an arbitrary complex vector $\xi$ with an appropriate dimension,
\begin{eqnarray}
& & \!\!\! \tilde{M}(\lambda,i)\tilde{V}_{2}^{[iv]}(\lambda,i)\xi \nonumber\\
&=& \!\!\! \tilde{U}(\lambda,i)
\!\left[\!\begin{array}{c}
{\rm\bf diag}\left\{\left.\tilde{\alpha}^{[j]}(\lambda)\right|_{j=1}^{\overline{\rm\bf m}(i)}\right\}  \\
0  \end{array}\!\right]\!\tilde{V}_{1}(\lambda,i)\tilde{V}_{2}^{[iv]}(\lambda,i)\xi \nonumber \\
&=& \!\!\! 0
\label{eqn:a31}
\end{eqnarray}

This is a contradiction with the assumption that the matrix pencil $\tilde{M}(\lambda,i)$ is of FNCR. Hence, $\overline{\rm\bf m}(i)={\rm\bf m}(i)$. This means that when the matrix pencil $\tilde{M}(\lambda,i)$ is of FNCR, it is column rank deficient only at a finite number of $\lambda\in {\cal C}$. Particularly, let ${\rm\bf\Lambda}_{1}(i)$ denote the set of the complex numbers at which the matrix pencil $\tilde{M}(\lambda,i)$ is column rank deficient. Then
\begin{equation}
{\rm\bf\Lambda}_{1}(i)=\bigcup_{j=1}^{{\rm\bf m}(i)}\left\{\;\lambda \; \left|\; \tilde{\alpha}^{[j]}(\lambda)=0,\;\;\lambda\in{\cal C} \right. \right\}
\label{eqn:a32}
\end{equation}

Let ${\rm\bf\Lambda}_{2}(i)$ denote the set of the complex numbers at which the matrix pencil $H_{\xi^{[i]}_{{\rm\bf H}}}(\lambda)$ is singular. Then from Lemma \ref{lemma:4}, this set also consists of only finitely many elements. On the other hand, Lemma \ref{lemma:4} also reveals that the matrix pencils $K_{\xi^{[i]}_{\rm\bf K}(j)}(\lambda)$ with $j=1, 2,\cdots, \zeta^{[i]}_{{\rm\bf K}}$ are not of FCR only at $\lambda=0$, while all the matrix pencils $N_{\xi^{[i]}_{\rm\bf N}(j)}(\lambda)$ with $j=1,2,\cdots,\zeta^{[i]}_{{\rm\bf N}}$ and $J_{\xi^{[i]}_{\rm\bf J}(j)}(\lambda)$ with $j=1,2,\cdots,\zeta^{[i]}_{{\rm\bf J}}$ are of FCR at each $\lambda\in{\cal C}$.

The above arguments show that if the matrix pencil $\tilde{M}(\lambda,i)$ is of FNCR, then for each
\begin{equation}
\lambda\in {\cal C}\left/\left\{{\rm\bf\Lambda}_{1}(i) \bigcup  {\rm\bf\Lambda}_{2}(i) \bigcup \{0\}\right\}\right.
\label{eqn:a33}
\end{equation}
all the matrix pencils $H_{\xi^{[i]}_{{\rm\bf H}}}(\lambda)$, $K_{\xi^{[i]}_{\rm\bf K}(j)}(\lambda)|_{j=1}^{\zeta^{[i]}_{{\rm\bf K}}}$, $N_{\xi^{[i]}_{\rm\bf N}(j)}(\lambda)|_{j=1}^{\zeta^{[i]}_{{\rm\bf N}}}$ and $J_{\xi^{[i]}_{\rm\bf J}(j)}(\lambda)|_{j=1}^{\zeta^{[i]}_{{\rm\bf J}}}$, as well as the matrix pencil $\tilde{M}(\lambda,i)$, are of FCR. As both the set ${\rm\bf\Lambda}_{1}(i)$ and the set ${\rm\bf\Lambda}_{2}(i)$ have only finitely many elements, the set ${\cal C}\left/\left\{{\rm\bf\Lambda}_{1}(i) \bigcup  {\rm\bf\Lambda}_{1}(i) \bigcup \{0\}\right\}\right.$ is not empty. Hence, the existence of the desirable $\lambda$ is guaranteed.

From Equation (\ref{eqn:a24}) and Lemma \ref{lemma:2}, as well as the block diagonal structure of the matrix pencil $K(\lambda,i)$, it can be further declared that at every $\lambda$ satisfying Equation (\ref{eqn:a33}), the matrix pencil  $\overline{M}(\lambda,i)$ is of FCR, also.

This completes the proof.   \hspace{\fill}$\Diamond$

\vspace{0.5cm}
\hspace*{-0.45cm}{\rm\bf Proof of Theorem \ref{theorem:4}:} Note that the requirement that the matrix pencil $\tilde{M}(\lambda,i)$ is of FNCR is equivalent to that the matrix pencil $\tilde{M}(-\lambda,i)$ is of FNCR. On the other hand, from Lemma \ref{lemma:4}, we know that for each $j=1,2,\cdots,\zeta^{[i]}_{{\rm\bf L}}$ and for an arbitrary $\lambda\in {\cal C}$,
\begin{equation}
{\rm\bf null}\left(\!\! L_{\xi^{[i]}_{\rm\bf L}(j)}(-\lambda)\!\!\right)=\left\{a_{j}{\rm\bf col}\!\left\{\left.1,\;\lambda^{k}\right|_{k=1}^{\xi^{[i]}_{\rm\bf L}(j)}\right\},\; a_{j}\in{\cal C}\right\}
\label{eqn:a34}
\end{equation}
Hence, for an arbitrary $\alpha$ satisfying
\begin{displaymath}
{\rm\bf diag}\!\left\{\!L_{\xi^{[i]}_{\rm\bf L}(j)}(\lambda)|_{j=1}^{\zeta^{[i]}_{{\rm\bf L}}}\right\}\alpha=0
\end{displaymath}
there certainly exist some complex numbers, denote them by $a_{1}$, $a_{2}$, $\cdots$ and $a_{\zeta^{[i]}_{{\rm\bf L}}}$, such that
\begin{equation}
\alpha
={\rm\bf col}\!\left\{\left.
a_{j} {\rm\bf col}\!\left\{\left.1,\;\lambda^{k}\right|_{k=1}^{\xi^{[i]}_{\rm\bf L}(j)}\right\}\right|_{j=1}^{\zeta^{[i]}_{{\rm\bf L}}}\right\}
\label{eqn:a35}
\end{equation}

On the contrary, direct matrix multiplications show that every vector $\alpha$ having an expression of Equation (\ref{eqn:a35}) belongs to the null space of the matrix ${\rm\bf diag}\!\left\{\!L_{\xi^{[i]}_{\rm\bf L}(j)}(\lambda)|_{j=1}^{\zeta^{[i]}_{{\rm\bf L}}}\right\}$.

Denote the vector ${\rm\bf col}\!\left\{\left. a_{j}\right|_{j=1}^{\zeta^{[i]}_{{\rm\bf L}}}\right\}$ by $a$. From the above observations, it is straightforward to prove that the matrix pencil $\tilde{M}(\lambda,i)$ is of FNCR, if and only if there exists a $\lambda\in {\cal C}$, such that
\begin{eqnarray}
& &\!\!\!\! \left[ P(i)\Theta(i)-\Pi(i)\right]{\rm\bf col}\!\left\{\left.
a_{j} {\rm\bf col}\!\left\{\left.1,\;\lambda^{k}\right|_{k=1}^{\xi^{[i]}_{\rm\bf L}(j)}\right\}\right|_{j=1}^{\zeta^{[i]}_{{\rm\bf L}}}\right\} \nonumber\\
&=& \!\!\!\!\left[P(i)\Theta(\lambda,i)-\Pi(\lambda,i)\right] a \nonumber \\
&\neq&\!\!\!\! 0
\label{eqn:a36}
\end{eqnarray}
for arbitrary complex numbers $a_{1}$, $a_{2}$, $\cdots$ and $a_{\zeta^{[i]}_{{\rm\bf L}}}$ that are not simultaneously equal to zero, which is equivalent to that $a\neq 0$. Therefore, the last inequality of Equation (\ref{eqn:a36}) exactly means that the MVP $P(i)\Theta(\lambda,i)-\Pi(\lambda,i)$ is of FNCR.

On the other hand, from the definition of the integer $\xi^{[i]}_{\rm\bf L}$, as well as those of the matrices $\overline{\Theta}_{j}(i)|_{j=1}^{\zeta^{[i]}_{{\rm\bf L}}}$ and $\overline{\Pi}_{j}(i)|_{j=1}^{\zeta^{[i]}_{{\rm\bf L}}}$, it is obvious that
\begin{eqnarray}
& &\!\!\!\! \left[ P(i)\Theta(i)-\Pi(i)\right]{\rm\bf col}\!\left\{\left.
a_{j} {\rm\bf col}\!\left\{\left.1,\;\lambda^{k}\right|_{k=1}^{\xi^{[i]}_{\rm\bf L}(j)}\right\}\right|_{j=1}^{\zeta^{[i]}_{{\rm\bf L}}}\right\} \nonumber\\
&=& \!\!\!\!\left\{\!\sum_{j=1}^{\zeta^{[i]}_{{\rm\bf L}}}\! a_{j}\! \left[P(i)\overline{\Theta}_{j}(i)-\overline{\Pi}_{j}(i)\right] \!\!\right\}
{\rm\bf col}\!\left\{\!\left.1,\;\lambda^{k}\right|_{k=1}^{\xi^{[i]}_{\rm\bf L}} \!\right\}
\label{eqn:a37}
\end{eqnarray}
It can therefore be declared that if the matrix pencil $\tilde{M}(\lambda,i)$ is of FNCR, then for every $a\neq 0$,
\begin{equation}
\sum_{j=1}^{\zeta^{[i]}_{{\rm\bf L}}}\! a_{j}\! \left[P(i)\overline{\Theta}_{j}(i)-\overline{\Pi}_{j}(i)\right] \neq 0
\label{eqn:a38}
\end{equation}

Note that
\begin{equation}
{\rm\bf vec}\!\left(\!\sum_{j=1}^{\zeta^{[i]}_{{\rm\bf L}}}\! a_{j}\! \left[P(i)\overline{\Theta}_{j}(i)-\overline{\Pi}_{j}(i)\right]\!\right)
=\Gamma(i)a
\label{eqn:a39}
\end{equation}
The inequality of Equation (\ref{eqn:a38}) implies that the matrix $\Gamma(i)$ is of FCR.

This completes the proof.   \hspace{\fill}$\Diamond$

\small



\begin{thebibliography}{60}
\bibitem{adm2020}{\rm F. {Anstett-Collin}, L. {Denis-Vidal} and G. $\rm Mill\acute{e}rioux$, "A priori identifiability: An overview on definitions and approaches", {\it Annual Reviews in Control}, Vol.50, pp.139$\sim$149, 2020.}

\bibitem{bv1988}{\rm T. Beelen and P. {Van Dooren}, "An improved algorithm for the computation of {Kronecker's} canonical form of a singular pencil", {\it Linear Algebra and Applications}, Vol.105, pp.9$\sim$65, 1988.}

\bibitem{cpakj2017}{\rm J. F. Carvalho, S. Pequito, A. P. Aguiar, S. Kar and K. H. Johansson, "Composability and controllability of structural linear time-invariant systems: distributed verification", {\it Automatica}, Vol.78, pp.123$\sim$134, 2017.}

\bibitem{cw2020}{\textcolor{black}{\rm V. Chetty and S. Warnick, "Meanings and applications of structure in networks of dynamic systems", {\it In} S. Roy and S. K. Das (Ed.) {\it Principles of Cyber-Physical Systems: An Interdisciplinary Approach}, Cambridge University Press, Cambridge, United Kingdom, 2020.}}

\bibitem{cw2017}{\textcolor{black}{\rm V. Chetty S. Warnick, "Necessary and sufficient conditions for identifiability of interconnected subsystems", {\it Proceedings of the 56th Conference on Decision and Control}, pp.5790$\sim$5795, Melbourne, Australia, 2017.}}


\bibitem{Gantmacher1959}{\rm F. R. Gantmacher, {\it The Theory of Matrices}, Chelsea, New York, USA, 1959.}


\bibitem{hgb2019}{\rm J. M. Hendrickx, M. Gevers and A. S. Bazanella, "Identifiability of dynamical networks with partial node measurements", {\it IEEE Transactions on Automatic Control}, Vol.64, No.6, pp.2240$\sim$2253, 2019.}

\bibitem{hj1991}{\rm R. A. Horn and C. R. Johnson, {\it Topics in Matrix Analysis}, Cambridge University Press, Cambridge, UK, 1991.}

\bibitem{htw2009}{\rm Y. F. Huang, I. Tienda-Luna and Y.F. Wang, "Reverse engineering gene regulatory networks", {\it IEEE Signal Processing Magazine}, Vol.26, No.1, pp.76$\sim$97, 2009.}

\bibitem{it2017}{\rm S. Iwata and M. Takamatsu, "On the {Kronecker} canonical form of singular mixed matrix pencils", {\it SIAM Journal on Control and Optimization}, Vol.55, No.3, pp.2134$\sim$2150, 2017.}

\bibitem{Kailath1980}{\rm T. Kailath, {\it Linear Systems}, Prentice Hall, Englewood Cliffs, New Jersey, USA, 1980.}

\bibitem{ptt2019}{\rm D. Patil, P. Tesi and S.Trenn, "Indiscernible topological variations in {DAE} networks", {\it Automatica}, Vol.101, pp.280$\sim$289, 2019.}

\bibitem{pka2016}{\rm S. Pequito, S. Kar and A. P. Aguiar, "A framework for structural input/output and control configuration selection in large-scale systems", {\it IEEE Transactions on Automatic Control}, Vol.61, No.2, pp.303$\sim$318, 2016.}

\bibitem{pmssaxcas2010}{\rm R. J. Prill, D. Marbach, J. Saez-Rodriguez, P. K. Sorger, L. G. Alexopoulos, X. W. Xue, N. D. Clarke, G. Altan-Bonnet and G.~Stolovitzky, "Towards a rigorous assessment of systems biology models: the dream3 challenges", {\it PLoS ONE}, Vol.5, No.2, e9202, 2010.}

\bibitem{Siljak1978}{\rm D. D. Siljak, {\it Large-scale Dynamic Systems: Stability and Structure}, North-Holland Books, New York, USA, 1978.}

\bibitem{scl2015}{\rm T. H. Summers, F. L. Cortesi and J. Lygeros, "On submodularity and controllability in complex dynamical networks", {\it  IEEE Transactions on Control of Network Systems}, Vol.3, No.1, pp.91$\sim$101, 2016.}

\bibitem{vdhb2013}{\rm P. M. J. {Van den Hof}, A. Dankers, P. S. C. Heuberger and X. Bombois, "Identification of dynamic models in complex networks with prediction error methods-basic methods for consistent module estimates", {\it Automatica}, Vol.49, No.10, pp.2294$\sim$3006, 2013.}

\bibitem{sbkkmpr2011}{\rm J. H. {Van Schuppen}, O. Boutin, P. L. Kempker, J. Komenda, T. Masopust, N. Pambakian and A. C. M. Ran, "Control of distributed systems: tutorial and overview", {\it European Journal of Control}, Vol.17, No.5-6, pp.579$\sim$602, 2011.}



\bibitem{vtc2021}{\textcolor{black}{\rm H. J. {Van Waarde}, P. Tesi and M. K. Camlibel, "Topology identification of heterogeneous networks: Identifiability and
reconstruction", {\it Automatica}, Vol. 123, 109331, 2021.}}


\bibitem{wvd2018}{\rm H. H. M. Weerts, P. M. J. {Van den Hof} and A. Dankers, "Identifiability of linear dynamic networks", {\it Automatica}, Vol.89, pp.247$\sim$258, 2018.}

\bibitem{wrmd2005}{\textcolor{black}{\rm J. C. Willems, P. Rapisarda, I. Markovsky and B. L. {DeMoor}, "A note on persistency of excitation", {\it Systems \& Control Letters}, Vol.54, No.4, pp.325$\sim$329, 2005.}}

\bibitem{xz2014}{\rm J. Xiong and T. Zhou, "Structure identification for gene regulatory networks via linearization and robust state estimation", {\it Automatica}, Vol.50, pp.2765$\sim$2776, 2014.}

\bibitem{zdg1996}{\rm K. M. Zhou, J. C. Doyle and K. Glover, {\it Robust and Optimal Control}, Prentice Hall, Upper Saddle River, New Jersey, USA, 1996.}

\bibitem{Zhou2015}{\rm T. Zhou, "On the controllability and observability of networked dynamic systems", {\it Automatica}, Vol.52, pp.63$\sim$75, 2015.}

\bibitem{zyl2018}{\rm T. Zhou, K. Y. You and T. Li, {\it Estimation and Control of Large Scale Networked Systems}, Elsevier, Butterworth-Heinemann, Oxford, UK, 2018.}

\bibitem{zz2020}{\rm T. Zhou and Y.Y. Zhou, "Affine dependence of network controllability/observability on its subsystem parameters and connections", {\it IEEE Transactions on Systems, Man, and Cybernetics: Systems}, doi:10.1109/TSMC.2020.3030765, 2020.}
\end{thebibliography}
\end{document}